\documentclass[reprint,aps,nofootinbib,superscriptaddress]{revtex4-2}
\usepackage[utf8]{inputenc}
\usepackage{amssymb,amsfonts,amsmath,bm}
\usepackage{graphics,graphicx}
\usepackage{hyperref}

\usepackage{braket}
\usepackage{mathtools}
\usepackage{float}
\usepackage{enumitem}

\usepackage{xcolor} 

\newcommand{\e}{{\textrm e}} 

\hypersetup{
    colorlinks = true,
    citecolor= black,
    linkcolor=black,
    urlcolor=cyan,
}

\newcommand{\GE}{\mathcal{G}_{\text{eff}}}

\newcommand{\GEinf}{\mathcal{G}^{\infty}_{\text{eff}}}

\begin{document}
	
	\title{Like-charge attraction at short distances in a charge-asymmetric two-dimensional two-component plasma: Exact results}
	
	\author{Lucas Varela}
	\affiliation{Universidad de los Andes, Bogot\'a, Colombia}
	\affiliation{Universit\'e Paris-Saclay, CNRS, LPTMS, 91405, Orsay, France.}	
	\author{Gabriel T\'ellez}
	\affiliation{Universidad de los Andes, Bogot\'a, Colombia}
	
	\begin{abstract}
		We determine exactly the short-distance effective potential between two ``guest'' charges immersed in a
		two-dimensional two-component charge-asymmetric plasma composed of positively ($q_1 = +1$) and negatively ($q_2 = -1/2$) charged point particles. The result is valid over the whole regime of stability, where the Coulombic coupling (dimensionless inverse temperature) $\beta <4$. At high Coulombic coupling $\beta>2$, this model features like-charge attraction. Also, there cannot be repulsion between opposite-charges at short-distances, at variance with large-distance interactions.
	\end{abstract}
	
	\maketitle		
	\makeatletter
	\let\toc@pre\relax
	\let\toc@post\relax
	\makeatother 

\section{Introduction}

 The study of soft matter covers a broad assortment of systems: polymers, foams, emulsions, liquid- and solid-aerosols, suspensions, etc \cite{Overbeek,hunterbook,jones2002soft}. 
 The interest in these systems is twofold: 
 a) they have a wealth of interesting and mostly counter-intuitive phenomena such as like-charge attraction \cite{LEVIN1999,Varenna}, charge reversal, self-assembly, electroosmosis \cite{Palaia2020,telles2021}, etc; 
 b) they are featured in different fields that range from material science (e.g.  cohesion in concrete \cite{pellenq2004,Ioannidou2016}) to biology \cite{Holm2001,Levin2002,andelman2006} (e.g. formation of DNA condensates \cite{bloomfield1996} and membrane dynamics \cite{Caffrey}). 
 Even though they may display widely different behaviors, they are rooted in three common characteristics: high responsiveness to thermal fluctuations, featuring two or more length scales (usually microscopic and mesoscopic) and the presence of strong collective effects. 
 Accounting for these properties implies dealing with a plethora of difficulties that make these systems hard to treat both numerically and analytically, even within simple models.
 
A key ingredient in soft matter models is long-range interactions, namely Coulomb electrostatic forces. 
In many-body systems, this alone may lead to a phenomenon know as like-charge attraction. Indeed, 
whereas two like-charges in vacuum will always repel, in the presence of an electrolyte they may attract. 
First evidence of this phenomenon dates back to the 1980s, where Monte Carlo simulations \cite{Guldbrand1984} and integral equations \cite{Kjellander1984} revealed that strongly-like-charged surfaces may attract through the mediation of counterions.  
Since then, there has been a number of results using simulations \cite{Moreira2002MC,dosSantos2018} and approximate analytic calculations \cite{Moreira2000,Naji2005,samaj2018}. However, exact results remain rare, which is where this work expects to contribute.   

\begin{figure}[htp]
	\centering
	\includegraphics[width=0.49\textwidth]{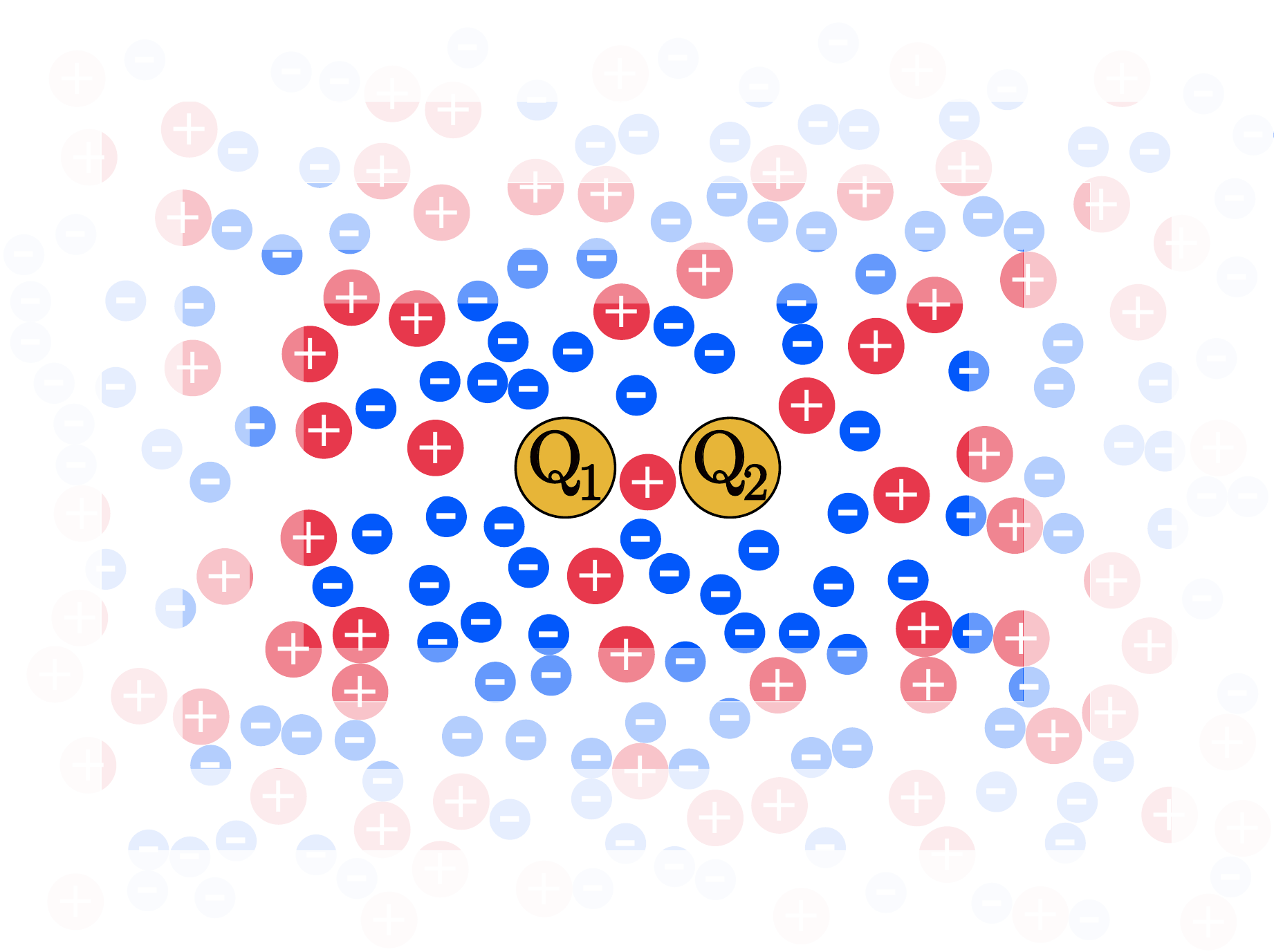}%
	\caption{Sketch of the charge-asymmetric two-dimensional two-component plasma with two guest charges: $Q_1$ and $Q_2$. The charges of the cations  (red) and anions (blue) are fixed to $+1$ and $-1/2$, whereas $Q_k (k=1,2)$ are studied for all possible values in the stability regime. The plasma occupies all $\mathbb{R}^2$.}%
	\label{fig:2DTCP_sketch}%
\end{figure}

Herein is reported the existence of like-charge attraction at short distances, in a classical two-dimensional (2D) Coulomb gas, from an exact calculation. We consider a system made 
 cations and anions, with charge strengths $q_1=1$ and $q_2=-1/2$  interacting via the 2D Coulomb potential. This potential, of logarithmic form, implies the existence of a regime where charges are stable against collapse, at variance with the `true' 3D potential $1/r$, which does require to include quantum mechanics to treat short-distance interactions \cite{lieb1976}.  Two `guest' charges are immersed in a gas, as envisioned in Fig.~\ref{fig:2DTCP_sketch}, and
the short-distance effective potential is determined. These two charges may attract for high enough coupling and charge strength, due to a mechanism which resembles the 1D one particle phenomenon \cite{Trizac2018}: an opposite-charge to the guest like-charges is shared by them, due to its impossibility to screen them both simultaneously. 

Low-dimensional models are suitable for a variety of treatments that have led to exact solutions, for one-dimensional \cite{lenard,lenard2,prager,baxter,Dean2009,frydel2019} and two-dimensional \cite{Jancovici1981,Hansen1985,Minnhagen_1987,cornu1989,samaj2003} Coulomb gases. For two-dimensional two-component plasmas in the stability regime, the system features a scale free potential which allows to compute the following exact equation of state
\begin{equation}
    \beta P = n\Big(1 + \frac{q_1q_2}{4} \beta \Big)
\end{equation}
where $P$ is the pressure and $\beta$ is the dimensionless inverse temperature. Furthermore, quantities like the internal energy, specific heat and temperature functions can be obtained exactly for some special cases by mapping onto an equivalent field theory. Indeed, these systems
are in correspondence with integrable 2D field theories but for some specific charges they correspond to extensively studied cases. For a charge-symmetric ($|q_1|=|q_2|$) system, the grand partition function can be mapped to the quantum sine-Gordon model with a conformal normalization of the cosine-field \cite{samaj2000}.  Moreover, this field theory can also be mapped to the massive Thirring model \cite{coleman1975}. In the quantum sine-Gordon model, the many-body correlation functions are expressed in terms of the expected values of the primary fields of the theory. 
These connections allow to compute the
pair correlation functions and effective potentials, at
short \cite{tellez_2005} and large \cite{samaj2000} distances. This allowed to show the symmetric case cannot feature like-charge attraction, opposite-charge repulsion
and overcharging \cite{samaj2005,Tellez2006Jstat}. Moreover, it was possible to show this is also true beyond the stability regime, by adding a hardcore interaction to avoid collapse \cite{samaj2006}. On the contrary, these phenomena are present in the special charge-asymmetric case $|q_1|/|q_2|= 2$. This system is equivalent to the complex Bullough-Dodd model, and analogously to the symmetric case, this connection allows to obtain the aforementioned thermodynamic functions \cite{samaj2003,tellez_2005}. This allows to show there is like-charge attraction, opposite-charge repulsion and overcharging at long distances \cite{tellez2006EPL}. The present work shows that there is like-charge attraction at short distances in the charge-asymmetric 2D two-component plasma, further distinguishing it from its symmetric counterpart. 

The paper is organized as follows. 
The short distance behavior of effective potential between two charges immersed in a two-dimensional two-component plasma is obtained analytically in Section \ref{sec:2DTCP}. 
This quantity allows to determine whether these two particles attract or repel in the presence of the many-body interactions with the plasma. 
Section \ref{sec:LCA} finds that there may be like-charge attraction between two negatively charged particles. 
The conditions for this phenomena are obtained and compared with the known results for the counterpart large-distance behavior~\cite{tellez2006EPL}. 

\section{The two-component plasma and the Complex Bullough–Dodd model}\label{sec:2DTCP}

Consider the charge-asymmetric two-dimensional two-component plasma (2D TCP) which consists of point-like cations and anions with respective charges $q_1=+1$ and $q_2 =- 1/\mathcal{Q}$ ($\mathcal{Q}>0$). 
This Coulomb gas is confined to an infinite 2D space and its constituents interact through the pair Coulomb potential $v$, given by
\begin{equation}
v(\mathbf{r}) = -\ln (r/r_0),
\label{2D_v}
\end{equation}
where $\mathbf{r}\in \mathbb{R}^2$, $r = |\mathbf r|$ and $v$ is the solution to the $2\text D$ Poisson equation ($\Delta v(\mathbf{r}) = -2\pi \delta(\mathbf{r})$). 
The gauge term $r_0$ determines the zero-energy reference distance of the Coulomb potential, which without loss of generality is set to unity.


Herein, the thermodynamics of the charge-asymmetric 2D TCP are worked out in the grand canonical ensemble, which features the grand partition function $\Xi$ given by 	
\begin{equation}
\Xi = \sum_{N_1,N_2=0}^{\infty} \int \prod_{j=1}^{N_1} d^2u_j z_1(\mathbf{u}_j)\prod_{j=1}^{N_2} d^2v_j  z_2(\mathbf{v}_j) \frac{\e^{-\beta H_{N_1,N_2}}}{N_1!N_2!},
\label{Xi_def}
\end{equation}
where $z_{\sigma}(\mathbf{u})$ ($\sigma=1,2$) are the position dependent fugacities for the cations and anions of the plasma and $\beta$ is the reduced dimensionless inverse temperature, also known as the Coulomb coupling. 
We assume that $\beta< 2\mathcal{Q}$, which is the so-called stability regime in which cations and anions interact without collapsing. The case $\beta\geq 2\mathcal{Q}$ requires the inclusion of a short-range repulsive force (e.g. hard-core potential). 
The dimensionless energy $H_{N_1,N_2}$ corresponds to a system with $N_{1}$ and $N_2$ particles of charge $q_1$ and $q_2$ respectively, given by
\begin{equation}
H_{N_1,N_2} = \sum_{j<k} q_{\sigma_j} q_{\sigma_k} 	v(|\mathbf{r}_j -\mathbf{r}_k|).
\end{equation}

The statistics of the 2D TCP is equivalent to a $2\text D$ Euclidean theory \cite{Minnhagen_1987}.
In particular, for the case $\mathcal{Q}=1$ (symmetric) and $\mathcal{Q}=2$ the Coulomb gas is connected to field theories that have been studied extensively: the quantum sine-Gordon \cite{samaj2000} and complex Bullough-Dodd \cite{samaj2003} models respectively.  In this paper we focus on the charge-asymmetric case $q_1 = 1$ and $q_2 = -1/2$ ($\mathcal{Q}=2$).
For this purpose, we summarize the main results known for this particular case, and we refer the reader to \cite{samaj2003} for a complete discussion and derivation. 

\subsection{Grand Partition Function}
First, the Boltzmann factor in the right-hand side of Eq.~\eqref{Xi_def} is re-expressed using the Hubbard–Stratonovich transformation and afterwards integrated by parts. The resulting expression is then used in Eq.~\eqref{Xi_def} to cast the grand partition function as  
\begin{equation}
\Xi = \frac{\int \mathcal{D}\phi \,\exp\left(-S[z_1,z_2]\right)}{\int \mathcal{D}\phi \,\exp\left(-S[0,0]\right)}
\label{Xi},
\end{equation}
where $\phi(\mathbf{r})$ is a real scalar field, $\int \mathcal{D}\phi$ is the functional integration over this field and $S$ is the action given by
\begin{equation}
S[z_1,z_2] = \int d^2r \Big[\frac{(\nabla\phi)^2}{16\pi} - z_1(\mathbf{r})\,\e^{ib\phi}-z_2(\mathbf{r})\,\e^{-i(b/2)\phi}\Big],
\label{action}
\end{equation}
where $b^2 = \beta/4$ and $\beta$ is the Coulombic coupling. With the field representation the multi-particle densities are related to the field averages, and in particular for the one- and two-body densities we have
\begin{align}
n_{\sigma} &= z_{\sigma} \braket{\e^{ibq_{\sigma}\phi}}, \\
n^{(2)}_{\sigma\sigma'}(|\mathbf{r}-\mathbf{r'}|) &= z_{\sigma}z_{\sigma'} \braket{\e^{ibq_{\sigma}\phi(\mathbf{r})}\e^{ibq_{\sigma'}\phi(\mathbf{r'})}},
\label{n2def} 	
\end{align}
where $\braket{\cdots}$ is the average with respect to  $S$ (Eq.~\eqref{action}). For the previous relation to be in correspondence with the charge-asymmetric ${+2}/{-1}$ 2D TCP, the fugacities  $z_{\sigma}$ ($\sigma=1,2$) have to be renormalized by the divergent self energy terms $\exp[\beta v(0)q_{\sigma}^2/2]$ and using the short-distance normalization 
\begin{equation}
\braket{\e^{ibq\phi(\mathbf{r})}\e^{ibq'\phi(\mathbf{r'})}} \sim |\mathbf{r}-\mathbf{r'}|^{\beta qq'} \braket{\e^{i(q+q')b\phi}}, \quad |\mathbf{r}-\mathbf{r'}| \to 0,
\label{short-distance-normalization}
\end{equation}	
where $\beta$ is assumed to be small enough~\cite{samaj2003}. The action $S$ equipped with the short-distance normalization forms a conformal field theory known as the complex Bullough–Dodd model \cite{FATEEV1998}, also known as the Zhiber–Mikhailov–Shabat model, which belongs to the affine Toda theories.

In \cite{Hansen1985}, the short-distance behavior of the pair distribution function was computed for an arbitrary charge-asymmetry in the canonical ensemble. For our purposes, we give summon their result in the case $\mathcal{Q}=2$, which is given by
\begin{equation}
    \begin{split}
    n_{q_1 q_1}(r) &\underset{r\to 0}{\sim} \begin{cases}
    r^{\beta} & \beta <2\\
    r^{2k-(4-k(9-k)/2)\beta/4} & \frac{8}{4-k+1}<\beta<\frac{8}{4-k}
    \end{cases}\\
        n_{q_2 q_2}(r) &\underset{r\to 0}{\sim} \begin{cases}
    r^{\beta/4} & \beta <2\\
    r^{2-3\beta/4} & 2<\beta<4
    \end{cases}
    \end{split}
    \label{hansen_paircorr}
\end{equation}
where $k\in \{1,2\}$. In section \ref{sec:short-distance} we give the exact behavior as $r\to0$ and in doing so, we recover the same $r$-dependence featured in Eq.~\eqref{hansen_paircorr}, for a grand canonical ensemble.

\subsection{The interaction potential between two external charges and the operator-product-expansion}

Consider two external point charges immersed in the plasma (see Fig.~\ref{fig:2DTCP_sketch}): $Q_1$ at the origin and $Q_2$ at $\mathbf{r}$. To avoid the collapse of the guest charges
with oppositely charged particles from the plasma, we suppose that ${-2}<\beta Q_{\sigma} < 4$ ($\sigma=1,2$).
We are interested in the effective potential between the guest 1,2-charges, which is defined by	
\begin{equation}
G_{Q_1Q_2}(r) = \mu^{\text{ex}}_{Q_1Q_2} (r) - \mu^{\text{ex}}_{Q_1} -\mu^{\text{ex}}_{Q_2},
\label{eff_pot_def}
\end{equation}
where $\mu_Q^{\text{ex}}$ is the excess chemical potential which is defined as the work required to move a charge $Q$ from infinity into the bulk of the plasma. Similarly, $\mu^{\text{ex}}_{Q_1Q_2}$ is defined as the work done to bring two guest charges $Q_1$ and $Q_2$ from infinity into the bulk of the gas at a distance $r$ apart. In \cite{samaj2005} the following expressions for the excess potentials where derived: 
\begin{equation}
\begin{split}
\exp(\beta \mu_Q^{\text{ex}}) &=  \frac{\Xi[Q]}{\Xi},
\\
\exp\left(\beta \mu^{\text{ex}}_{Q_1Q_2} (r)\right) &=  \frac{\Xi[Q_1,0;Q_2,\mathbf{r}]}{\Xi},
\end{split} 
\end{equation}
where $\Xi[Q]$ is the grand partition
function of the plasma in the presence of a guest-charge $Q$, and
$\Xi[Q_1,0;Q_2,\mathbf{r}]$ of a
charge $Q_1$ at the origin and $Q_2$ at $\mathbf{r}$. We remind that $\Xi$ is the grand partition function of the plasma without external charges (Eq.~\eqref{Xi_def}). 
Then, inserting the previous results into the effective potential (Eq.~\eqref{eff_pot_def}), it was found in \cite{samaj2005} the following expression for $G_{Q_1Q_2}(r)$ in terms of the grand partition function
\begin{equation}
\e^{-\beta G_{Q_1Q_2}(r)} = \frac{\Xi[Q_1,0;Q_2,\mathbf{r}]/\Xi}{(\Xi[Q_1]/\Xi)(\Xi[Q_2]/\Xi)}
\label{exp_eff_pot_xi}.
\end{equation}

Note that the position functional dependence of the effective potential is solely given by $\Xi[Q_1,0;Q_2,\mathbf{r}]$. This grand partition function can be expanded as the sum of terms that correspond to canonical systems where there are two guest charges and a finite number of plasma particles. For concreteness sake we list the first terms that appear in terms of total number of particles. The first term consists of two guest charges and  zero plasma particles. This contributes to $G_{Q_1Q_2}(r)$ with a bare Coulomb interaction $-Q_1 Q_2 \ln r$. Next, we have the two guest charges and we either add one cation or anion, for a total 3 particles and we continue increasing the number of plasma charges.   
 
For the purpose of illustrating how the terms of the previous expansion behave, we compute one of the terms featuring the guest charges $Q_{\sigma}$ $(\sigma=1,2)$, and a single plasma particle ${+1}$. We would like to know the effective potential between the charges $Q_1$ and $Q_2$ as a function of their separation $r$. The positions of $Q_1$ and $Q_2$ are fixed at the origin and $\mathbf{r}$ respectively, while the cation can be anywhere in the available space. The partition function for this system is
\begin{equation}
Z_{1}[Q_1,0;Q_2,\mathbf{r}] = r^{\beta Q_1 Q_2}\int d\bm{r'}\;({r'})^{\beta Q_1}|\bm{r} - \bm{r'}|^{\beta Q_2  },
\label{Z1_1}
\end{equation}
where $\bm{r'}$ is the cation's position. We can rescale $\bm{\widehat{r}'} = \bm{r'}/r$ in Eq.~\eqref{Z1_1}, and in doing so we obtain 
\begin{equation}
\begin{split}
Z_{1}=r^{\beta (Q_1 Q_2+Q_1+Q_2)+2}
\int d\bm{\widehat{r}'}\;  (\widehat{r}')^{\beta Q_1}|\bm{\widehat{r}'}-\bm{1}|^{\beta Q_2 },
\end{split}\label{Z1_2}
\end{equation}
where we have used the rotational symmetry to change the term $\bm{\widehat{r}} \to \bm{1}$ in the integral. With the previous change of variables we withdraw the $r$ dependence out of integral and obtain the explicit functional form of the partition function in terms of the guest-charge separation. Note that the integral in Eq.~\eqref{Z1_2} is convergent
if $\beta Q_1>-2$ (i.e. stability condition on $Q_1$) and $\beta(Q_1+Q_2)<-2$. The threshold in the later condition will play an important role in the effective potential expansion: it marks a transition of the the dominant term in the short-distance behavior. For $\beta(Q_1+Q_2)> -2$ this integral has an analytic continuation (discussed later) and the contribution associated to Eq.~\eqref{Z1_2} recedes its dominant status.

Equation \eqref{exp_eff_pot_xi} is obtained by summing over terms which successively add plasma particles to the system they represent. In what follows we will find an explicit and formal expression for short-distance limit $r\to 0$, which will later be used to identify the dominating interaction. For this purpose the effective potential is cast in terms of the field theory, which has the following equality that was derived in \cite{samaj2005}
\begin{equation}
\e^{-\beta G_{Q_1Q_2}(r)} = \frac{\braket{\e^{ibQ_1\phi(\mathbf{0})}\e^{ibQ_2\phi(\mathbf{r})}} }{ \braket{\e^{ibQ_1\phi}} \braket{\e^{ibQ_2\phi}} }.
\label{exp_eff_pot}
\end{equation}
The expectation fields for the complex Bullough–Dodd model have been studied extensively and their analytic expression has been derived in \cite{FATEEV1998}
\begin{equation}
\begin{split}
&\braket{\e^{ia\phi}} =\\ &\quad\left[ \frac{z_2\Gamma(1+b^2)\Gamma(\frac{4-b^2}{4})}{z_12^{\frac{b^2}{2}}\Gamma(1-b^2)\Gamma(\frac{4+b^2}{4})}\right]^{\frac{2a}{3b}} 
\left[\frac{\Gamma(\frac{3+\xi}{3})\Gamma(\frac{1+\xi}{2})}{m^{-1}4^{\frac{1}{3}}\sqrt{3}\Gamma(\frac{1}{3})}\right]^{2a^2-ab}
\\ 
&\times 
\exp\Big\{\int_{\mathbb{R}^{+}} \frac{dt}{t} \Big[\frac{\sinh((2-b^2)t)\text{csch}(2t)\Psi(t,a)}{\sinh(3(2-b^2)t) \sinh(b^2t)} - \frac{2a^2}{\e^{2t}}\Big]\Big\},
\end{split}
\end{equation}
where $\xi=b^2/(2-b^2)$ and
\begin{equation}
\begin{split}
&\Psi(t,a) = - \sinh(2abt)\big\{\sinh([4-b^2-2ab]t)\\
&-\sinh([2-2b^2+2ab]t)	+\sinh([2-b^2-2ab]t)\\
&-\sinh([2-b^2+2ab]t) -\sinh([2+b^2-2ab]t)
\big\},
\end{split}
\end{equation}
and
\begin{equation}
\begin{split}
m = \frac{2\sqrt{3} \,\Gamma(\frac{1}{3})}{\Gamma(\frac{3-\xi}{3})\Gamma(\frac{1+\xi}{3})} \left[\sqrt{\frac{z_1 \Gamma(1-b^2)}{\Gamma(b^2)/\pi}\;}\frac{z_2 \Gamma(\frac{4-b^2}{4})}{\Gamma(\frac{b^2}{4})/2\pi}\right]^{(1+\xi)/3}
,
\end{split}
\label{mass_breather}
\end{equation}
where the term $m$ corresponds to the mass of the lightest particle, a 1-breather, featured in the complex Bullough-Dodd  model.

We are interested in the effective potential between the guest charges when they are at short range. For this purpose, we use the short-distance operator-product-expansion of $\braket{\e^{ibQ_1\phi(\mathbf{0})}\e^{ibQ_2\phi(\mathbf{r})}} $, which has been computed in \cite{BASEILHAC}: 
\begin{equation}
\begin{split}
\e^{ibQ_1\phi(\mathbf{0})}\e^{ibQ_2\phi(\mathbf{r})}  = \sum_{n=0}^{\infty} \Big\{ C_{Q_1Q_2}^{n,0} (r) \e^{ib(Q_1+Q_2+n) \phi} + \cdots\Big\}\\
+\sum_{n=1}^{\infty} \Big\{ {C'}_{Q_1Q_2}^{n,0} (r) \e^{ib(Q_1+Q_2-n/2)\phi} + \cdots\Big\} \\+ \sum_{n=1}^{\infty} \Big\{ D_{Q_1Q_2}^{n,0} (r) \e^{ib(Q_1+Q_2+n-1/2)\phi}  + \cdots\Big\}.
\end{split}
\label{ope}
\end{equation}
The previous equation is the kind of expansion we were looking for, which will be shown later to contain the terms that correspond to systems with a finite number of plasma particles. For this purpose we require the coefficients in Eq.~\eqref{ope}, that were computed in \cite{BASEILHAC}
\begin{equation}
\begin{split}
C_{Q_1Q_2}^{n,0} (r) &= z_1^n f_{Q_1Q_2}^{n,0}(z_1z_2^2 r^{6-3b^2})  \\
&\times  r^{4Q_1Q_2b^2 + 4nb^2(Q_1+Q_2) + 2n(1-b^2) + 2n^2b^2}, 
\\
{C'}_{Q_1Q_2}^{n,0} (r) &=z_2^n {f'}_{Q_1Q_2}^{n,0}(z_1z_2^2 r^{6-3b^2}) 
\\
&\times    r^{4Q_1Q_2b^2 - 2nb^2(Q_1+Q_2) + 2n(1-b^2/4) + n^2b^2/2}, \\
D_{Q_1Q_2}^{n,0} (r) &=z_2z_1^n g_{Q_1Q_2}^{n,0}(z_1z_2^2 r^{6-3b^2})  \\  
\times &  r^{4Q_1Q_2b^2 + (4n-2)b^2(Q_1+Q_2) + 2n(1-2b^2) +2+ 2n^2b^2}.
\label{coefficients}
\end{split}
\end{equation}

For each function $h\in \{f,f',g\}$ there exists a power series expansion:
\begin{equation}
h_{Q_1Q_2}^{n,0}(t) = \sum_{k=0}^{\infty} h_{k}^{n,0} (Q_1,Q_2)t^k,
\end{equation}
where the leading terms of the previous expansion are 
\begin{equation}
\begin{split}
f_{0}^{n,0} (Q_1,Q_2) &= j_n(Q_1, Q_2,1) \quad(\text{for } n\neq 0)\\
{f'}_{0}^{n,0} (Q_1,Q_2) &=j_n(-Q_1/2,-Q_2/2,1/4)\\
g_{0}^{n,0} (Q_1,Q_2) &= \mathcal{F}_{n,1} (Q_1, Q_2,1), 
\end{split}
\end{equation}
and

\begin{equation}
\begin{split}
j_n(a_1, a_2,\rho) =& \frac{1}{n!} \int \prod_{k=1}^{n} d^2x_k \prod_{k=1}^{n} |x_k|^{a_1\beta}|1-x_k|^{a_2\beta} 
 \\
 &\times\prod_{k<p}^{n}|x_k-x_p|^{\rho\beta},
\label{jn}\end{split}
\end{equation}
\begin{equation}
\begin{split}
&\mathcal{F}_{n,m}(a_1, a_2,\rho) = \frac{1}{n!m!} \int \prod_{k=1}^{n} d^2x_k \int  \prod_{l=1}^{m} d^2y_l  
\\
&\times\prod_{k=1}^{n}|x_k|^{a_1\beta}|1-x_k|^{a_2\beta}   \prod_{k<p}^{n}|x_k-x_p|^{\rho\beta}\prod_{l=1}^{m} |y_l|^{- a_1\beta/2}
\\
&\times |1-y_l|^{-a_2\beta/2}\prod_{l<q}^{m} |y_l-y_q|^{\rho\beta/4} \prod_{k,l}^{n,m}|x_k-y_l|^{-\rho\beta/2},
\label{Fnm}\end{split}
\end{equation}
Note that $j_n(Q_1, Q_2,1)$ is proportional to the configuration integral of a Coulomb gas, made of $n$ cations ($+1$) and two fixed guest charges: $Q_1$ at the origin and $Q_2$ at $\mathbf{1}$. Likewise, the term $j_n(-Q_1/2, -Q_2/2,1/4)$ is connected to an analogous system where cations are replaced by anions ($-1/2$). Finally, the function $\mathcal{F}_{n,m}(Q_1, Q_2,1)$ is related to a system with $n$ cations, $m$ anions and the two guest charges. 
Therefore, these configuration integrals correspond to a charge-asymmetric ${+2}/{-1}$ 2D TCP with finite amount of plasma particles (i.e. canonical ensemble) and two guest charges. These cases are of special interest in the discussion to follow since they appear in the short-distance expansion, which allows to identify the respective term they appear with as the interaction associated to one of the previously discussed $N$ body systems. 

The integral $j_n(a_1, a_2,\rho)$ is known as the complex Selberg integral, and it was independently studied in \cite{DOTSENKO,DOTSENKO2} and \cite{AOMOTO}, with the following outcome:
\begin{equation}
\begin{split}
j_n&(a_1,a_2,\rho) = \Big[\frac{\pi}{\gamma(\rho\beta/4)}\Big]^n \prod_{k=1}^{n}\gamma\Big(\frac{k\rho\beta}{4}\Big) 
\\ 
&\times \prod_{k=0}^{n-1}\gamma\Big(1+\frac{\beta[2a_1+k\rho]}{4}\Big)
\gamma\Big(1+\frac{\beta[2a_2+k\rho]}{4}\Big)
\\
&\times 
\gamma\big(-1-{\beta[2a_1+2a_2+(n-1+k)\rho]}/4\big),
\end{split}
\end{equation}
where $\gamma(x) = \Gamma(x)/\Gamma(1-x)$. 

Replacing Eq.~\eqref{ope} in Eq.~\eqref{exp_eff_pot} we obtain
\begin{equation}
\label{exp_eff_pot_expansion}
\begin{split}
&\e^{-\beta G_{Q_1Q_2}(r)} =  \frac{\braket{\e^{ib(Q_1+Q_2)\phi}} }{ \braket{\e^{ibQ_1\phi}} \braket{\e^{ibQ_2\phi}} }\, r^{\mathfrak{F}^{\text{bare}}}
\\
&\hspace{3.6cm}\times \big[1  +\mathcal{O}\big(r^{\min\{6-3b^2,4\}}\big)\big]
\\
&+\sum_{n=1}^{\infty}\frac{\braket{\e^{ib(Q_1+Q_2+n)\phi}} }{ \braket{\e^{ibQ_1\phi}} \braket{\e^{ibQ_2\phi}} }\, z_1^n \,j_n(Q_1,Q_2,1)\,r^{\mathfrak{F}^{\text c}_{n}} 
\\
&\hspace{4.5cm}\times\big[1 +\mathcal{O}\big(r^{6-3b^2}\big)\big]
\\
&+\sum_{n=1}^{\infty}\frac{\braket{\e^{ib(Q_1+Q_2-\frac{n}{2})\phi}} }{ \braket{\e^{ibQ_1\phi}} \braket{\e^{ibQ_2\phi}} }\, z_2^n \, j_n\Big(-\frac{Q_1}{2},-\frac{Q_2}{2},\frac{1}{4}\Big)\,r^{\mathfrak{F}^{\text a}_{n}} 
\\
&\hspace{4.5cm}\times\big[1 +\mathcal{O}\big(r^{6-3b^2}\big)\big]
\\
&+\sum_{n=1}^{\infty}\frac{\braket{\e^{ib[Q_1+Q_2+(n-\frac{1}{2})]\phi}} }{ \braket{\e^{ibQ_1\phi}} \braket{\e^{ibQ_2\phi}} }\, z_2\,z_1^n \,\mathcal{F}_{n,1}(Q_1,Q_2,1)\,r^{\mathfrak{F}^{\text{a,c}}_{1,n}} \\
&\hspace{4.5cm}\times\big[1 +\mathcal{O}\big(r^{6-3b^2}\big)\big],
\end{split}
\end{equation}
where 
\begin{equation}
\begin{split}
\mathfrak{F}^{\text{bare}} &= \beta Q_1Q_2,\\
\mathfrak{F}^{\text c}_{n} =& \beta Q_1Q_2 + \beta n(Q_1+Q_2) + n\Big(2-\frac{\beta}{2}\Big) + \frac{\beta n^2}{2},\\
\mathfrak{F}^{\text a}_{n} =& \beta Q_1Q_2 - \frac{\beta n}{2} (Q_1+Q_2) + n\Big(2-\frac{\beta}{8}\Big) + \frac{\beta n^2}{8},  \\
\mathfrak{F}^{\text{a,c}}_{1,n} =& \beta Q_1Q_2 + \beta \Big(n-\frac{1}{2}\Big)(Q_1+Q_2) + n(2-\beta) + \frac{\beta n^2}{2}\\
&+2.
\label{exponents}
\end{split}
\end{equation}
These exponents are defined for $n\geq 1$, as seen in Eq.~\eqref{exp_eff_pot_expansion}. Note that the smallest function in Eq.~\eqref{exponents} is in the dominant term of Eq.~\eqref{exp_eff_pot_expansion}, when $r\to 0$.

\begin{figure}[htp]
	\centering
	\includegraphics[width=0.49\textwidth]{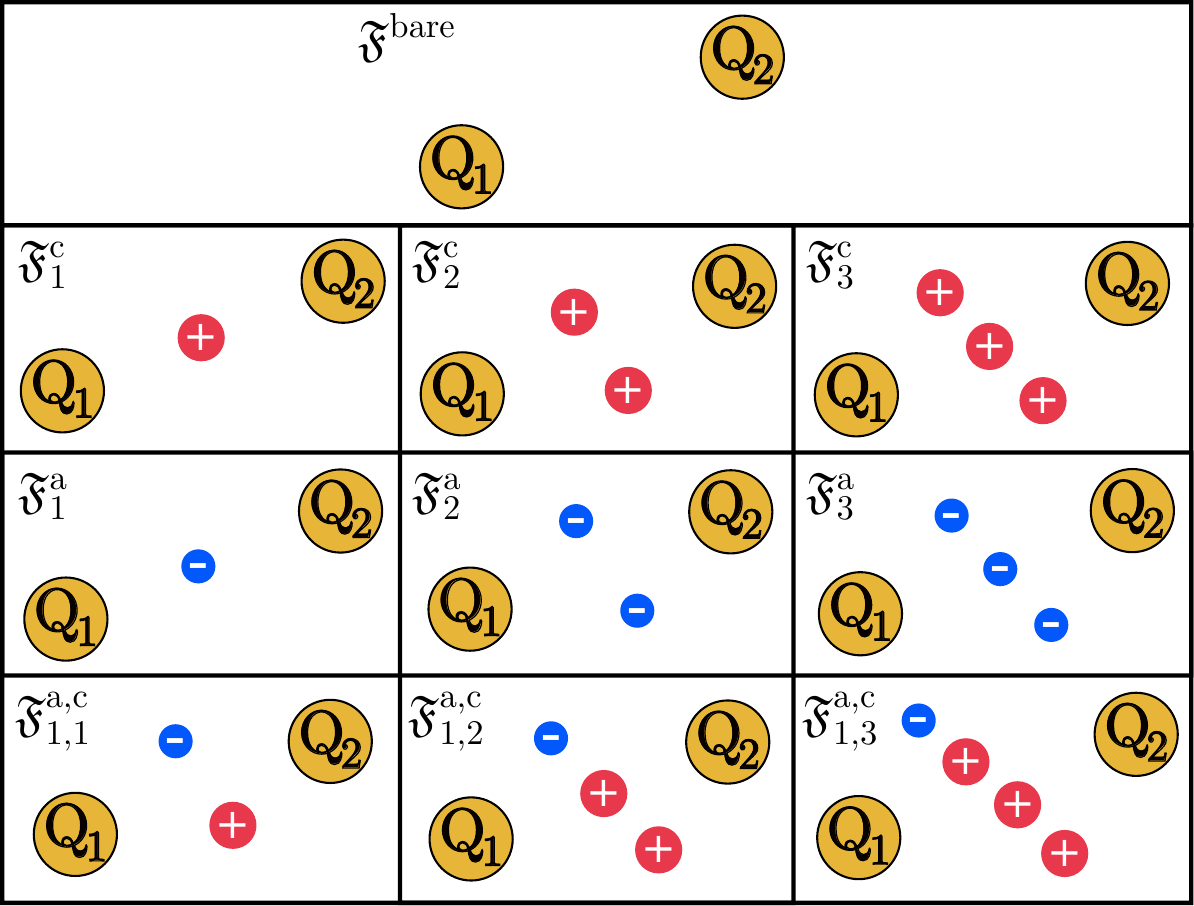}%
	\caption{Sketches of the systems associated to the configuration integrals $\{r^{\mathfrak{F}^{\text{bare}}},j_n r^{\mathfrak{F}^{\text c}_n}, j_n r^{\mathfrak{F}^{\text a}_n},\mathcal{F}_{n,1} r^{\mathfrak{F}_{1,n}^{{\text{a,c}}}}\}$, for $n=1,2,3$. Each cell depicts a system which identified by its respective $\mathfrak{F}$-function. The case ${\mathfrak{F}^{\text{bare}}}$ is for the bare guest charges.  The rest of the systems are made of the guest charges and they include the following plasma particles:  ${\mathfrak{F}^{\text c}_n}$ adds $n$ cations;  ${\mathfrak{F}^{\text a}_n}$ adds $n$ anions; $ {\mathfrak{F}_{1,n}^{{\text{a,c}}}}$ adds a single anion and $n$ cations.}%
	\label{fig:F_sketch}%
\end{figure}

Equation \eqref{exp_eff_pot_expansion} has the following physically interpretation: the rhs is a sum over terms associated to systems with a finite number of charges, in resemblance with a grand partition function. The terms $\{r^{\mathfrak{F}^{\text{bare}}},j_n r^{\mathfrak{F}^{\text c}_n}, j_n r^{\mathfrak{F}^{\text a}_n},\mathcal{F}_{n,1} r^{\mathfrak{F}_{1,n}^{{\text{a,c}}}}\}$ are configuration integrals of systems with a finite number of particles. Figure \ref{fig:F_sketch} depicts the systems associated to these configuration integrals. Note that, up to an additive constant,
the free energy of these systems is $-\mathfrak{F} \log r$, with the respective $\mathfrak{F} \in \{{\mathfrak{F}^{\text{bare}}},{\mathfrak{F}^{\text c}_n},  {\mathfrak{F}^{\text a}_n}, {\mathfrak{F}_{1,n}^{{\text{a,c}}}}\}$. Then, these $\mathfrak{F}$-functions are closely related to free energies. 
Note that, up to an additive constant, the effective potential $G_{Q_1Q_2}$ behaves as one of the aforementioned free energies, in the limit $r\to0$. The following section moves on to describe the hierarchy of the $\mathfrak{F}$-functions, which allows to determine the dominant term featured in the effective potential expansion at short-distances. 

\section{Short and large distance asymptotic potential}

In this section we compute the short-distance behavior for the effective potential. Then, we summon and briefly discuss the results found in \cite{tellez_2005}, where the large-distance behavior for $ G_{Q_1Q_2}$ was obtained analytically. This will allow to compare the two asymptotic results. 

\subsection{Short-distance asymptotic potential}
\label{sec:short-distance}

The dominant term of the short-distance effective potential (Eq.~\eqref{exp_eff_pot_expansion}) has
the power law with the minimum exponent. Therefore, to find this dominant interaction we determine which function in Eq.~\eqref{exponents} yields the minimum value, for a given set of parameters: guest charges $Q_1,Q_2$ and a coupling parameter $\beta$.
One way to proceed is by comparing these functions in three possible cases, based on guest-charge signs: positive $(Q_1,Q_2 >0)$, negative $(Q_1,Q_2 <0)$, and oppositely charged. This process is straightforward and it reveals that actually, the dominant term depends only on $\beta(Q_1+Q_2)$. We summarize the results in the following equation:   
\begin{widetext}
\begin{align}
\e^{-\beta G_{Q_1Q_2}(r)} \underset{r\to 0}{\sim}
\begin{dcases}
\frac{\braket{\e^{i\sqrt{\beta}(Q_1+Q_2-n/2)\phi/2}} }{ \braket{\e^{i\sqrt{\beta}Q_1\phi/2}} \braket{\e^{i\sqrt{\beta}Q_2\phi/2}} } z_2^n r^{\mathfrak{F}^{\text a}_{n}} j_n\Big(-\frac{Q_1}{2},-\frac{Q_2}{2},\frac{1}{4}\Big), &4+\frac{(n-1)\beta}{2} < \beta (Q_1 + Q_2) < 4+\frac{n\beta}{2},
\\
\frac{\braket{\e^{i\sqrt{\beta}(Q_1+Q_2)\phi/2}} }{ \braket{\e^{i\sqrt{\beta}Q_1\phi/2}} \braket{\e^{i\sqrt{\beta}Q_2\phi/2}} } r^{\mathfrak{F}^{\text{bare}}}, & -2 < \beta(Q_1+Q_2) < 4,
\\
\frac{\braket{\e^{i\sqrt{\beta}(Q_1+Q_2+n)\phi/2}} }{ \braket{\e^{i\sqrt{\beta}Q_1\phi/2}} \braket{\e^{i\sqrt{\beta}Q_2\phi/2}} } z_1^n r^{\mathfrak{F}^{\text{c}}_{n}} j_n\left(Q_1,Q_2,1\right), & -2 - n \beta < \beta(Q_1+Q_2) < -2 - (n-1) \beta, 
\end{dcases}
\label{exp_eff_pot_short}
\end{align}
\end{widetext}
where $n \geq 1$ is an integer and $\mathfrak{F}^{\text a}_{n},\mathfrak{F}^{\text{bare}}$ and $\mathfrak{F}^{\text c}_{n}$ are given by Eq.~\eqref{exponents}. This relation is valid provided that we are still in the region of stability of the system;
$-2<\beta Q_{\sigma} < 4$ ($\sigma=1,2$) and $\beta < 4$. The intervals in Eq.~\eqref{exp_eff_pot_short} can be understood as follows: at short distance, the guest-charges form a cluster with the minimum quantity of plasma particles necessary such that the net charge of the group satisfies the stability condition with both cations and anions. Namely, they form a cluster with minimum $n_c (n_a)$ cations (anions), such that $-2<\beta(Q_1+Q_2+n_c-n_a/2)<4$. In practice, we see that $n_c$ or/and $n_a$ is zero.

Note that if we take $Q$ as cation or anion, then $\exp(-\beta G_{Q Q})$ has the same $r$ dependence as the two body density. Then, it can be seen that Eq.~\eqref{hansen_paircorr} and Eq.~\eqref{exp_eff_pot_short} have the same position dependence.
The behavior of Eq.~\eqref{hansen_paircorr} holds for an arbitrary integer asymmetry ($|q_1/q_2| \in \mathbb{N}$). It has an identical interpretation to the present case: the effective interaction of two charges in the plasma is given by a cluster made of the two particles plus a few plasma charges (or none at all). Then, we surmise that at short distances, for both arbitrary guest-charges and charge-asymmetry of the plasma, they form the minimal cluster with a net charge that is stable against collapse with both cation and anions. Then, the effective potential has the same $r$-dependence as the cluster.

At short distances, the effective potential has the following functional form
\begin{equation}
\beta G_{Q_1Q_2} \sim -\mathcal{G}_{\text{eff}} \ln r,
\label{EGeff}
\end{equation}
where $\mathcal{G}_{\text{eff}}$ will be referred to as the interaction strength,
which from Eq.~\eqref{exp_eff_pot_short} is given by
\begin{equation}
\begin{split}
 \mathcal{G}_{\text{eff}}&(Q_1,Q_2;\beta) =\\
 &\begin{cases}
 \mathfrak{F}^{\text a}_{n}, &4+\frac{(n-1)\beta}{2} < \beta (Q_1 + Q_2) < 4+\frac{n\beta}{2},
 \\
 \mathfrak{F}^{\text{bare}}, &-2 < \beta(Q_1+Q_2) < 4,
 \\
 \mathfrak{F}^{\text c}_{n},  &-2 - n \beta < \beta(Q_1+Q_2) < -2 - (n-1) \beta. 
 \end{cases}  
 \label{Geff}
 \end{split}
\end{equation}
where $\mathfrak{F}^{\text a}_{n},\mathfrak{F}^{\text{bare}}$ and $\mathfrak{F}^{\text c}_{n}$ are given by Eq.~\eqref{exponents}.
The interaction strength between the two guest charges has a stellar role in the discussion that follows: its sign determines whether the the guest charges attract or repel. Hence, two like-charges attract when they feature a negative interaction strength. 

\begin{figure}[htp]
	\centering
	\includegraphics[width=0.49\textwidth]{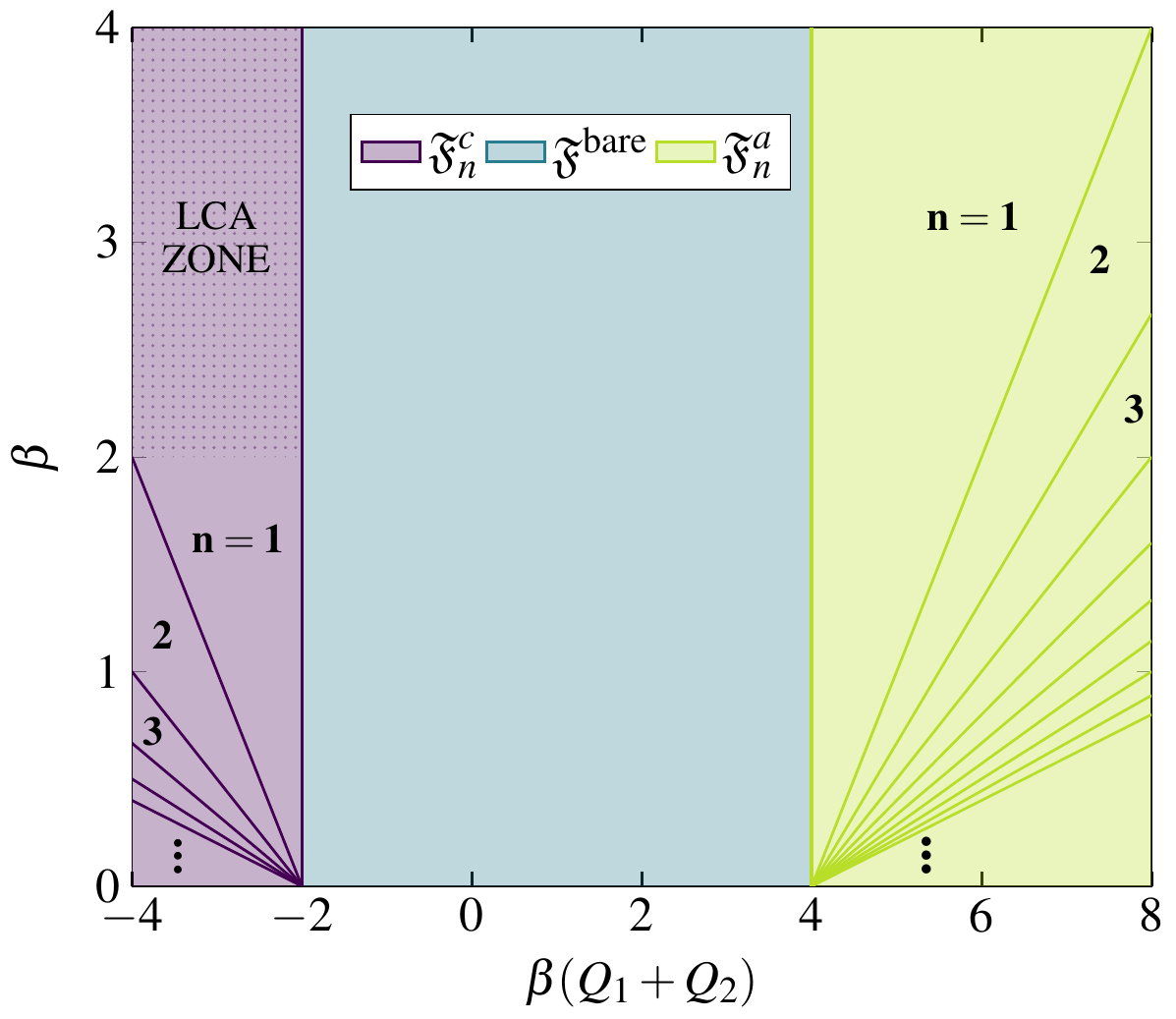}%
	\caption{Landscape of the interaction strength, which is the dominant power law coefficient in the exact short-distance expansion of the effective potential (Eq.~\eqref{exp_eff_pot_expansion}) between two guest charges ($Q_1$ and $Q_2$) immersed in a two-dimensional two-component plasma. The interaction strength (Eq.~\eqref{Geff}) has three main expressions which correspond to each panel, which from left to right are: $\mathfrak{F}^{\text{c}}_{n}$, $\mathfrak{F}^{\text{bare}}$ and  $\mathfrak{F}^{\text{a}}_{n}$ (Eq.~\eqref{coefficients}). The natural number $n$ is given for each sector. The plot shows that the functional form landscape of the interaction strength is determined by sum of charge of the two guest particles $Q_{1}+Q_2$, together with the  Coulomb coupling (inverse dimensionless temperature) $\beta$. However, we stress that $\mathfrak{F}^{\text{c}}_{n}$, $\mathfrak{F}^{\text{bare}}$ and  $\mathfrak{F}^{\text{a}}_{n}$ are functions of $Q_1,Q_2$ and $\beta$, not simply of $(Q_1+Q_2)$ and $\beta$. 
	LCA ZONE  is the the region where there may be like-charge attraction: $-4\beta (Q_1+Q_2)<-2$ and $\beta>2$. The entire stability regime is present: $0<\beta<4$ and $-4<\beta(Q_1+Q_2)<8$. 
	Note there are vertical dots at the bottom of the left and right panels, which indicate there is an infinity of sectors.}%
	\label{fig:F_landscape}%
\end{figure}

Figure \ref{fig:F_landscape} shows the function landscape $\mathfrak{F}$ for the interaction strength $\GE$. It features the entire stability regime, which is surrounded by the collapse zone. Within each panel the ruling coefficient is, from left to right, $\mathfrak{F}^{\text{c}}_{n}, \mathfrak{F}^{\text{bare}}$ and $\mathfrak{F}^{\text a}_{n}$. The subdivisions of the rectangles show the value of $n$ for the respective coefficient, which ranges from $0$ to $\infty$. 
We will see that the most interesting case is for $\mathfrak{F}^{\text{c}}_{1}$, where there is like-charge attraction.
The region where $\mathfrak{F}^{\text{bare}}$ is dominant, given by the interval $ -2 < \beta(Q_1+Q_2) < 4$, has a bare-charge interaction $\beta G_{Q_1Q_2} \sim -Q_1Q_2 \ln r$. This regime contains all the oppositely charged guest-charge cases without collapse and hence, they always have an attractive interaction, as expected.
Note that for plasma charges, the short-distance bare-charge interaction strength is consistent with the short-distance normalization Eq.~\eqref{short-distance-normalization}, where the connection between $\exp (-\beta G_{q q'})$ and $\braket{\e^{ibq\phi(\mathbf{r})}\e^{ibq'\phi(\mathbf{r'})}}$ is given by Eq.~\eqref{exp_eff_pot}. 

When $\beta(Q_1+Q_2)\not\in[-2,4]$, the coefficients $\mathfrak{F}_1^a$ or $\mathfrak{F}_1^b$ enter into the play. Notice that this regime corresponds to a situation of instability if both guest charges are seen as a single charge $Q_1+Q_2$. If $\beta(Q_1+Q_2)<-2$, a single charge $Q_1+Q_2$ will collapse with an ion of charge $+1$ of the plasma. This indicates the need to consider configurations with the two guest charges $Q_1$, $Q_2$ and an ion $+1$ of the plasma, and then the dominant term of $G_{Q_1Q_2}(r)$ will be given by $-\mathfrak{F}_1^c \ln r$ which corresponds to this configuration. Similarly, if $\beta(Q_1+Q_2)>4$, a single charge $Q_1+Q_2$ will collapse with an ion of charge $-1/2$ of the plasma. The relevant configuration here is the two guest charges $Q_1$, $Q_2$ and an ion $-1/2$ of the plasma, leading to $G_{Q_1Q_2}(r)\sim -\mathfrak{F}_1^a \ln r$.

\subsection{Large-distance asymptotic potential}

Equation \eqref{exp_eff_pot} also has a large-distance expansion $(r\to \infty)$. Using the form-factors of the exponential fields in the complex Bullough-Dodd model \cite{samaj2003,Smirnov1992}, it was found in \cite{tellez2006EPL} that the large-distance behavior for the effective potential is given by
\begin{equation}
   \beta G_{Q_1Q_2} \underset{r\to \infty}{\sim} \mathcal{G}_{\text{eff}}^{\infty}\, K_0 (mr), \label{long_range_eff_pot}
\end{equation}
where $K_0(x)$ is the modified Bessel function of order zero, $m$ is the same in Eq.~\eqref{mass_breather} and $\GEinf$ is the interaction strength at large distances given by
\begin{equation}
    \mathcal{G}_{\text{eff}}^{\infty} = \frac{8\sqrt{3\,}\,\prod_{\sigma=1}^2\sin\big(\frac{2\pi \beta Q_{\sigma}}{3(8-\beta)}\big)\cos\big(\frac{2\pi\beta Q_{\sigma}}{3(8-\beta)}-\frac{\pi (1+2\xi)}{6}\big)}{ \pi\,\e^{-\mathcal{I}}\, \sin\big(\frac{2\pi \xi}{3}\big)\sin\big(\frac{2\pi (1+\xi)}{3}\big)},
    \label{Geff_rinfty}
\end{equation}
where $\xi$ has the same definition used in the previous sections and
\begin{equation}
    \mathcal{I} = -4\int_0^{\infty} \frac{\cosh(\frac{t}{6}) \sinh(\frac{\xi t}{3})\sinh\big(\frac{(1+\xi)t}{3}\big)}{\sinh (t) \cosh(t/2)} \frac{dt}{t},
\end{equation}
and for $\beta <8/3$, which is the regime studied in \cite{tellez2006EPL}.  At large distances, $\GEinf$ plays the analogous role of $\GE$: two charges attract when $\GEinf$ is negative and repel when positive.  However, bewared that it only makes sense to compare the signs $\GE$ and $\GEinf$, not the magnitude.  Indeed, their magnitudes are only relevant when used in the complete expression for $G_{Q_1Q_2}$.
It was found in \cite{tellez2006EPL} that only negative charges can attract each other and furthermore, that the regime where this happens is
\begin{equation}
    \begin{split}
   \beta\,Q_{1,2} &< \beta - 4, 
   \\
    \beta - 4&< \beta\, Q_{2,1} < 0,
    \\
    \beta &>2.
    \end{split}
    \label{large_regime}
\end{equation}
It was also found in \cite{tellez2006EPL} that two oppositely charged particles can repel, provided that $Q_{1,2}>0$ and $Q_{2,1}<\beta - 4<0$.
%
\section{Like-charge attraction}
\label{sec:LCA}

In this section we give the regime where there is like-charge attraction between two guest charges immersed in the charge-asymmetric 2D TCP, at short distances. Namely, we give the region where $\mathcal{G}_{\text{eff}}$ is negative for like-charges.  We begin by considering two particular cases of interest: an interaction where at least one of the particles belongs to the plasma and then, between two identical guest charges $Q_1=Q_2$.   
We conclude with the conditions for like-charge attraction between two arbitrary guest charges, provided that they are within the stability regime. Throughout this section  short- and long-distance behaviors are compared. 

\subsection{Particular cases}

\subsubsection{Interactions involving plasma charges}

There are four possibilities which involve at least one plasma particle: cation-cation, anion-anion, cation-$Q_1$ and anion-$Q_1$ , where $-2<\beta Q_1<4$ is a guest charge. We begin by discussing the cation-cation case, where the interaction strength takes the form  of $\mathfrak{F}^{\text{bare}}$ and  $\mathfrak{F}^{\text a}_{n}$. The former is a bare-charge interaction, as seen in Eq.~\eqref{exponents}. Therefore,  when $\GE = \mathfrak{F}^{\text{bare}}$ there cannot be like-charge attraction. Besides, it is straightforward to show that, for this case, $\mathfrak{F}^{\text a}_{n}$ is always positive. This ensues a repulsive interaction and consequently, cations may never attract each other.

\begin{figure}[htp]
	\centering
	\includegraphics[width=0.49\textwidth]{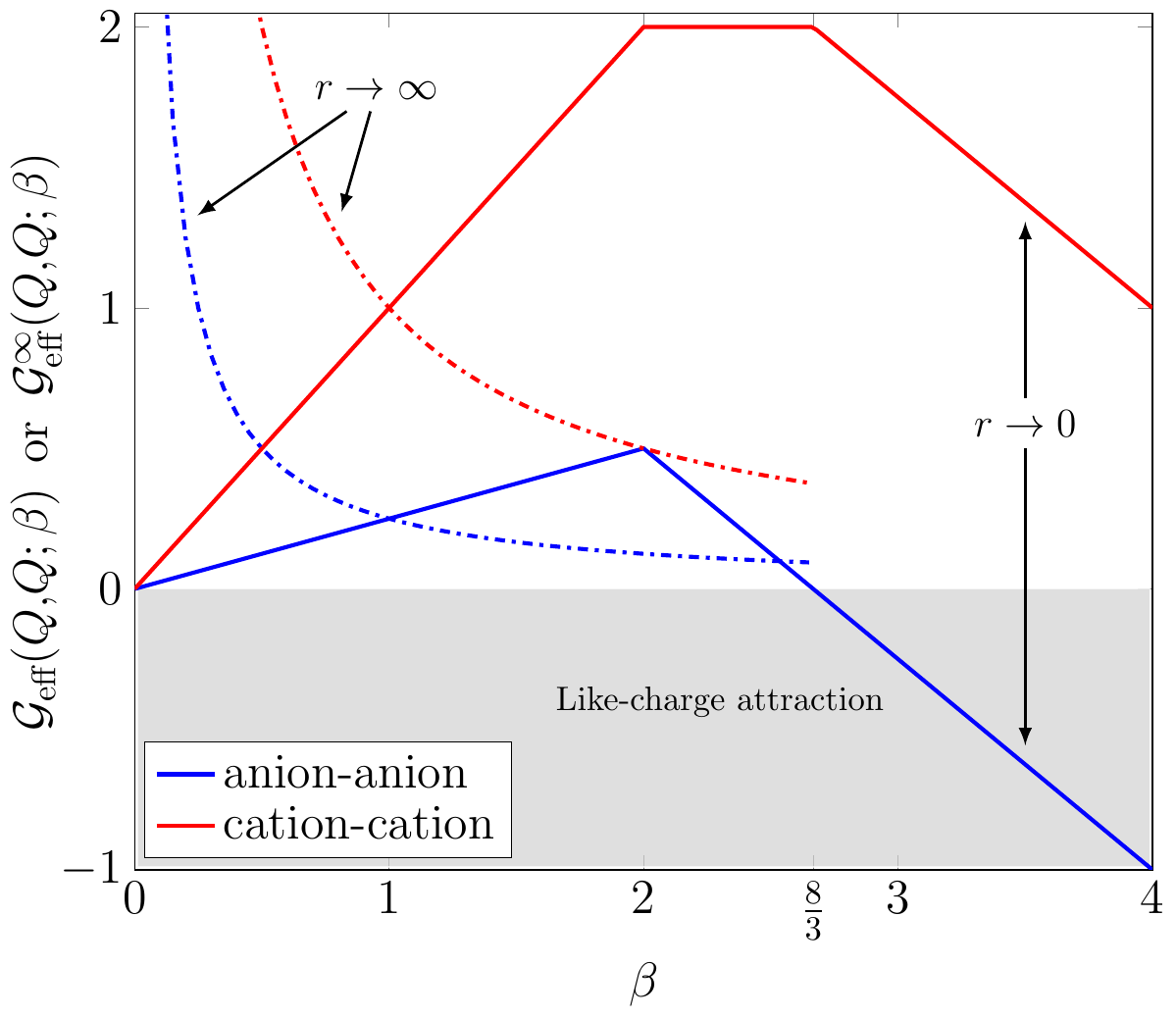}%
	\caption{Short (solid) and large (dotdashed) distance interaction strength  between two like-charged plasma particles: $Q = 1$ cations (red) and $Q=-1/2$ anions (blue). The anions feature like-charge attraction (curve within shaded region) for $\beta > 8/3$.  The large-distance interaction strength is extracted from \cite{tellez2006EPL}, where the results are valid for $\beta<8/3$.}%
	\label{fig:anions_cations}%
\end{figure}

Next, we consider the anion-anion interaction. The fact that both charges are negative entails that $\mathcal{G}_{\text{eff}}$ is either $\mathfrak{F}^{\text{bare}}$ or $\mathfrak{F}^{\text c}_{n}$, as seen in Eq.~\eqref{Geff}. We disregard the bare-charge interaction due to the aforementioned reasons and proceed to examine $\mathfrak{F}^{\text c}_{n}$.  It is straightforward to see that $\mathfrak{F}^{\text c}_{1}$ becomes negative at high Coulombic couplings, namely $\beta>8/3$. Hence, two anions indeed attract provided they are at small enough temperatures (large $\beta$). 
 Figure  \ref{fig:anions_cations} shows the cation-cation and anion-anion interaction strength as a function of the Coulomb coupling, at both short and large distances.
Note that the large-distance interaction strength is always positive for the known \cite{beta8d3} Coulomb couplings, at variance with the short-distance case. 

\begin{figure}[htp]
	\centering
	\includegraphics[width=0.49\textwidth]{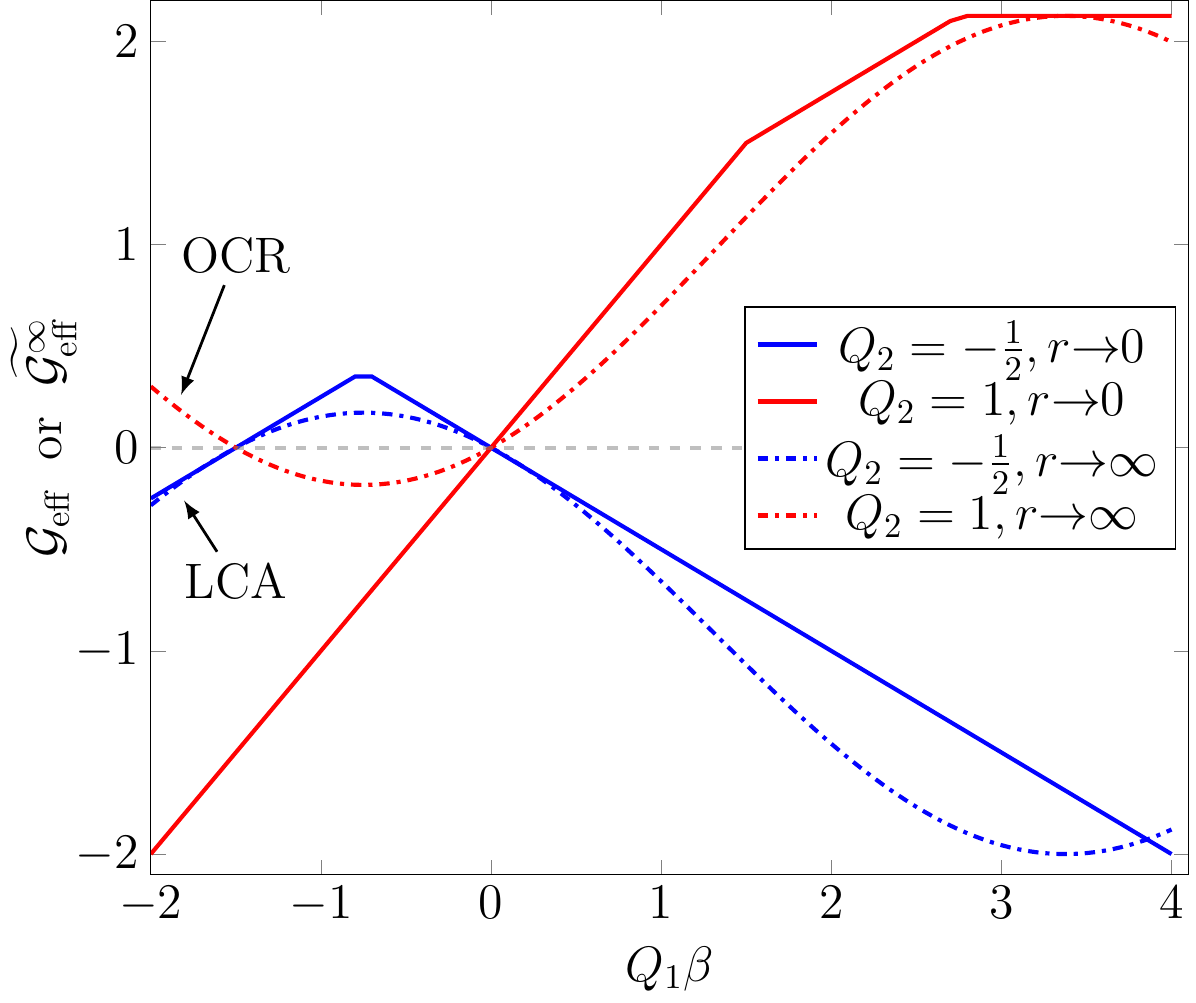}%
	\caption{Short (solid) and large (dotdashed) distance interaction strength between a guest-charge $Q_1$ and a plasma particle, cation (red)  and anion (blue), for a Coulombic coupling (inverse dimensionless temperature) $\beta =2.5$.  In the figure, LCA and OCR stand for like-charge attraction and opposite-charge repulsion respectively. 
	There is like-charge attraction between $Q_1$ and the anion when $\beta Q_1< -1.5$, for both short (Eq.~\eqref{LCA_R_anion_Q}) and large distances (Eq.~\eqref{large_regime}). 
	Contrarily, there is only opposite-charge repulsion at large distances, for the case of a cation interacting with $Q_1$.
	Note that this figure includes the complete stability interval for $Q_1$:  $-2<\beta Q_1 <4$. The large-distance interaction strength is extracted from \cite{tellez2006EPL}, and it is normalized so to accommodate $\GE$ and $\GEinf$ in the same plot \cite{normG}.}%
	\label{fig:Q_anions_cations}%
\end{figure}

We move on to examine the interaction of a guest-charge, which without loss of generality we call $Q_1$, with either a cation ($Q_2=1$) or anion ($Q_2=-1/2$). The former is governed by $\mathfrak{F}^{\text{bare}}$ and $\mathfrak{F}^{\text a}_{n}$, which as for the cation-cation case are non-negative. Hence, a cation in the charge-asymmetric 2D TCP will always repel  like charges. Contrarily, an anion may indeed  attract a like-charge. This follows from the term $\mathfrak{F}^{\text c}_{1}$, which is negative in the regime defined by the following inequalities:
\begin{equation}
\begin{split}
-2<\beta Q_1 < \beta -4 \quad\text{and}\quad
2<\beta<4,
\end{split}
\label{LCA_R_anion_Q}
\end{equation} 
where $\beta<4$ and $-2<\beta Q_1$ come from the stability condition. The remaining inequalities ensure that $\GE <0$.  Figure \ref{fig:Q_anions_cations} features the interaction strength for the cation-$Q_1$ and anion-$Q_1$ cases, at both short and large distances.
The anion case has the same attraction/repulsion regions for both distance asymptotics, whereas for the cation there is a major difference: at large distances the cation can repel with an opposite-charge,  at variance with short distances.
Note that in Fig.~\ref{fig:Q_anions_cations}, the interaction strength $\GEinf$ is normalized. This allows to evidence the sign changes of $\GE$ and $\GEinf$ in the same figure. Recall that these quantities stem from different functional expressions and therefore cannot by compared by their magnitudes, as previously discussed.

\subsubsection{Identical guest charges}

\begin{figure}[htp]
	\centering
	\includegraphics[width=0.49\textwidth]{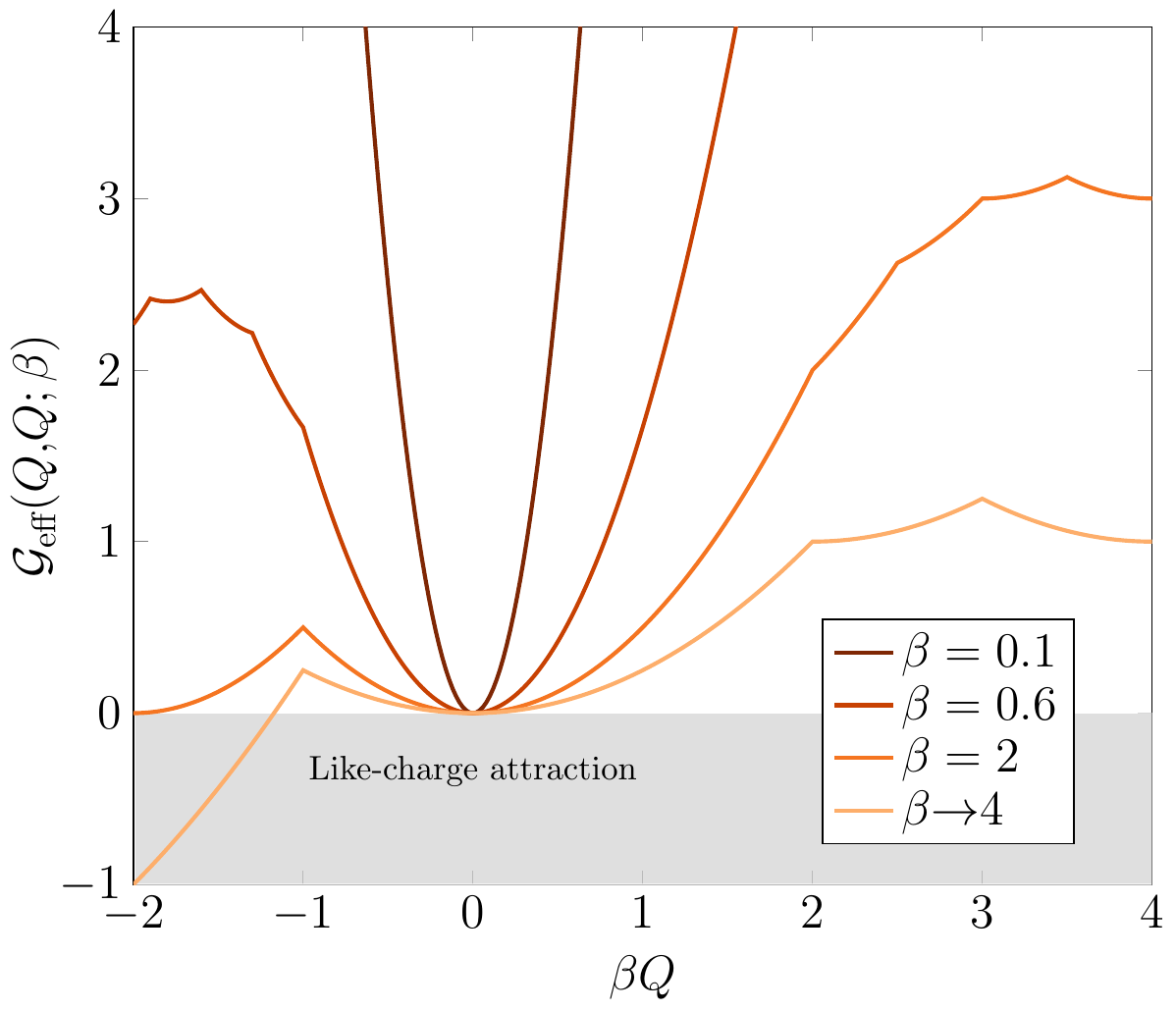}%
	\caption{Short-distance interaction strength $\GE$ (Eq.~\eqref{Geff}) between two identical guest charges $Q_1=Q_2=Q$, as a function of $\beta Q$. For large enough Coulomb couplings ($\beta>2$), there may be like-charge attraction (curve within the shaded region), as seen here for $\beta \to 4$.}%
	\label{fig:identicalcharges}%
\end{figure}

We move to search for like-charge attraction between two identical guest charges: $Q=Q_1=Q_2$. When the interaction strength is $\mathfrak{F}^{\text{bare}}$ there cannot be like-charge attraction, since this term corresponds to a  bare-charge interaction. Then, the only possibility is that  $\mathfrak{F}^{\text c}_{n}$ and/or $\mathfrak{F}^{\text a}_{n}$ are negative, in the respective intervals where they are dominant (see Eq.~\eqref{Geff}). It turns out that $\mathfrak{F}^{\text a}_{n}$ is always positive. Hence, two identical positively charged particles may never attract.  However,  $\mathfrak{F}^{\text c}_{n}$ does become negative in the following interval
\begin{equation}
\begin{split}
-2<\beta Q <  \sqrt{\beta-2}- \beta \quad\text{and}\quad
2<\beta <4,
\end{split}
\label{LCA_identic_charge_interval}
\end{equation}
where the lower and upper bounds of $\beta Q$ and $\beta$ respectively are the stability requirements. Therefore two identical negative charges attract, provided that Eq.~\eqref{LCA_identic_charge_interval} is satisfied. 
The interaction strength is featured in Fig.~\ref{fig:identicalcharges}, for a few Coulomb coupling numbers. 
For small $\beta<2$ we see that $\GE >0$, whereas for large values it becomes negative and therefore there is like-charge attraction. In contrast, at large distances this may never happen: from Eq.~\eqref{Geff_rinfty} we know that the interaction strength is positive since it goes like $\GEinf(Q,Q;\beta) \sim  (Q_{\text{eff}}(Q))^2$, where $Q_{\text{eff}}(Q)$ is some real function of $Q$. 

\subsection{Arbitrary guest charges \texorpdfstring{$Q_1$, $Q_2$}{}}

This section proceeds to determine the conditions (i.e. charge and temperature regime) for like-charge attraction 
between two guest particles $Q_{\sigma} (\sigma=1,2)$, at short distances.  We find that negative like-charges may attract each other whereas positive ones cannot. 
Let us begin by considering the latter. For two positive charges the interaction strength is ruled by either  $\mathfrak{F}^{\text{bare}}$ or $\mathfrak{F}^{\text a}_{n}$.
When $\mathfrak{F}^{\text{bare}}$ dominates, the bare-charge interaction cannot lead to like-charge attraction and in Appendix \ref{appx_LCA_2DTCP}, we show that neither does $\mathfrak{F}^{\text a}_{n}$. Hence, positive charges in the charge-asymmetric 2D TCP cannot attract at short distances.     

We move on to consider two negative charges $Q_{\sigma} <0$ $(\sigma=1,2)$. When $\mathfrak{F}^{\text{bare}}$ dominates, the bare-charge interaction cannot lead to like-charge attraction and therefore we move to examine $\mathfrak{F}^{\text c}_{n}$ in the interval $-4<\beta (Q_1+Q_2) <-2$ (see Fig.~\ref{fig:F_landscape}).
In Appendix \ref{appx_LCA_2DTCP} we show that $\mathfrak{F}^{\text c}_{n}$ is positive for $n>1$. However, $n=1$ is special since $\mathfrak{F}^{\text c}_{1}$ is negative if the following inequality is satisfied:
\begin{equation}
\mathfrak{F}^{\text c}_{1}= \beta Q_1Q_2 - \beta(|Q_1|+|Q_2|) + 2  < 0.
\label{Fc1_neg}
\end{equation}
It can readily be seen that this inequality can only be satisfied if $|Q_1|< 1$ and $|Q_2|< 1$ simultaneously.  We now examine the dependence on the Coulomb coupling, for which we solve for $\beta$:
\begin{equation}
	\beta > \frac{2}{|Q_1|+|Q_2|-Q_1 Q_2},
	\label{b*}
\end{equation}
Note that the threshold for like-charge attraction to manifest is $\beta>2$.  To summarize, the like-charge attraction regime is defined by the following inequalities:
\begin{equation}
\begin{split}
\frac{2-\beta|Q_{2,1}|}{1-|Q_{2,1}|} <\beta|Q_{1,2}|  < \beta \quad\text{and}\quad
2<\beta <4.
\end{split}
\label{LCA_REGIME}
\end{equation}
Note that this regime is contained within the region where $\mathfrak{F}^{\text c}_{1}$ is the interaction strength (Eq.~\eqref{Geff}).

\begin{widetext}
	\begin{figure*}[htp] 
		\includegraphics[width=0.99\textwidth]{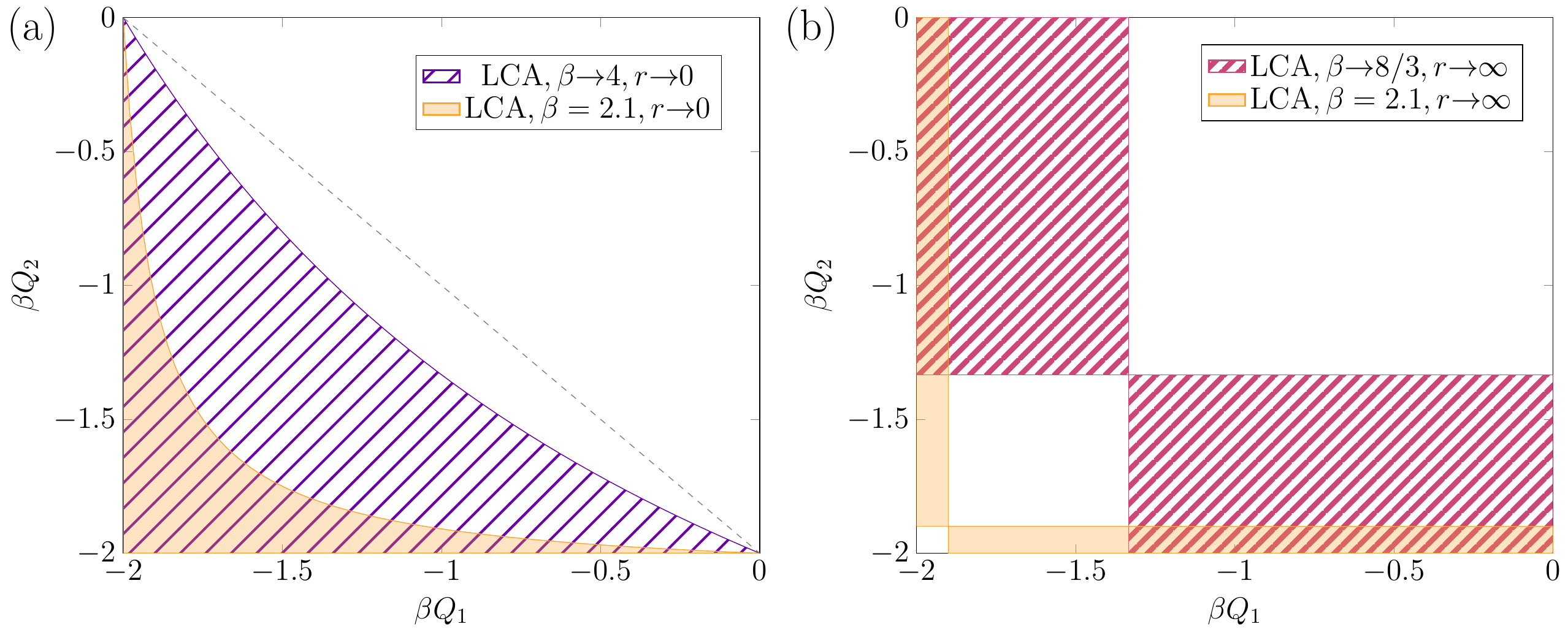}
		\caption{(a) Short-distance ($r\to 0$) like-charge attraction regimes for $\beta\to 4$ (stripes) and $\beta=2.1$ (orange filling). The threshold when  like-charge attraction may occur is $\beta>2$. Below the dashed line $\mathcal{G}_{\text{eff}} = \mathfrak{F}^{\text c}_{n}$ and above (within the plot domain)  $\mathcal{G}_{\text{eff}} =\mathfrak{F}^{\text{bare}}$, for any Coulomb coupling within the stability regime $(0<\beta<4)$. Note that the stability regime requires  $\beta Q_{1,2} > -2$.
			The limit $\beta \to 4$ refers to the asymptotic region as the Coulomb coupling approaches the collapse threshold, $4$.
			(b) Large-distance ($r\to \infty$) like-charge attraction regimes for $\beta\to 8/3$ (thick stripes) and $\beta=2.1$ (orange filling). The threshold when like-charge attraction may occur is $\beta>2$. 
			The large-distance behavior was obtained in \cite{tellez2006EPL} and the case $\beta \to 8/3$ refers to the asymptotic region at that Coulomb coupling number since the results found therein are valid for $\beta <8/3$.}
		\label{fig:LCA_regions}
	\end{figure*} 
	\end{widetext}

Figure $\ref{fig:LCA_regions}$ shows the regions where like charges attract, for a given Coulomb coupling at (a) short distances (Eq.~\eqref{LCA_REGIME}) and (b) large distances (Eq.~\eqref{large_regime}). By comparing these figures it can be seen the regimes where $\mathcal{G}_{\text{eff}}$ and $\mathcal{G}_{\text{eff}}^{\infty}$  feature like-charge attraction are considerably different.
Whereas two identical charges that are close together may attract (see Fig.~\ref{fig:identicalcharges} or \ref{fig:LCA_regions}a), they will always repel at large separations ($Q_1=Q_2$ is empty in Fig.~\ref{fig:LCA_regions}b). We had already witness this feature during the discussion of interactions among like-charged plasma particles.  Indeed we see that for a given Coulomb coupling, the like-charge regime for short distances does not contain its large-distance counterpart, and conversely. One characteristic they do share is the Coulomb coupling threshold for this phenomena: $\beta >2$.  
Besides finding the presence of like-charge attraction at short distances, we also showed that oppositely charged particles have a bare-charge interaction. Consequently, they will always attract at variance to the large-distance interactions.

So far, we have not discussed the mid-range distance behavior for the 2D TCP. Although we do not have results for that case, we can surmise what happens based on the limiting cases. For some guest charges and Coulombic coupling, there is like-charge attraction at both $r\to 0$ and $r\to\infty$. Then, this hints to the possibility that there is also attraction at mid-range distances, as opposed to having an effective force that changes sign twice.

We conclude with a remark on the effective interactions for the charge-symmetric 2D TCP, which is known for short \cite{tellez_2005} and large distances \cite{samaj2005}. In this system, the effective interaction between like-charges is always repulsive. Besides, oppositely charged particles always attract. Then, for the symmetric 2D TCP the common knowledge that like-charges repel and unlike-charges attract is true, whereas in its asymmetric counterpart this intuitive behavior may completely break down. 

%
%
%

\section{Conclusions}

We find that there may be like-charge attraction at short distances in the charge-asymmetric ${+2}/{-}1$ two-dimensional two-component plasma. More precisely, between negative charges (see Figs.~\ref{fig:anions_cations}-\ref{fig:identicalcharges}) and at high enough Coulomb coupling (i.e. small temperatures). Furthermore, we determine the charge and Coulomb coupling domain where this phenomena takes place (see Fig.~\ref{fig:LCA_regions}a).
Like-charge attraction is traced to a 3-body interaction, where a negative charge pairs with a plasma cation $(+1)$ to attract the other negative charge. 
This results are compared to the large distances behavior, which also features like-charge attraction (see Fig.~\ref{fig:Q_anions_cations}). However, the large-distance interaction may lead to opposite-charges to repel, a possibility that is absent at short distances.   
The short-distance result are in contrast to the symmetric two-dimensional two-component plasma, where like-charge attraction cannot happen at short distances (\cite{tellez_2005}). 

\section*{Acknowledgement}

We would like to thank E. Trizac and L. \v{S}amaj for useful discussions. This work was supported by an ECOS-Nord/Minciencias C18P01 action of Colombian and French cooperation. L.V. and G.T. acknowledge support from Fondo de Investigaciones, Facultad de Ciencias, Universidad de los Andes INV-2019-84-1825. L.V. acknowledges support from Action Doctorale Internationale (ADI 2018) de l'IDEX Universit\'e Paris-Saclay.

\appendix
\section{Analysis of  $\mathfrak{F}^a_n$ and $\mathfrak{F}^c_n$}
\label{appx_LCA_2DTCP}

This appendix shows that $\mathfrak{F}^c_n$ and $\mathfrak{F}^a_n$ are positive for $n>1$ and $n\geq 1$ respectively, in the intervals where they are associated with the interaction strength (Eq.~\eqref{Geff}). We begin by considering $\mathfrak{F}^c_n$, which is $\GE$ in the region that satisfies the following inequality: $-2-n\beta <\beta (Q_1+Q_2) <-2-(n-1) \beta$. 
 The former can be done by showing that $\mathfrak{F}^c_n$ is positive in a bigger simpler square region: $\mathcal{R}_{Q_1Q_2}^{c} = [-2/\beta,0]\times[-2/\beta,0]$. This set contains all the possible negative-charge values that satisfy the stability condition. We proceed to show that  the minimum of $\mathfrak{F}^c_n$ is positive in $\mathcal{R}_{Q_1Q_2}^{c}$, and consequently so does $\mathfrak{F}^c_n$. 
Since $\mathfrak{F}^c_n$ does not have critical points in $\mathcal{R}_{Q_1Q_2}^{c}$, the minimum lies in boundary of  $\mathcal{R}_{Q_1Q_2}^{c}$. It is straightforward to see that the minimum lies on the vertex points of the square boundary:
\begin{align}
\begin{split}
	\mathfrak{F}^{\text c}_{n}(Q_1=0,Q_2 = 0) &= n\Big(2-\frac{\beta}{2}\Big) + \frac{\beta n^2}{2},\\
	\mathfrak{F}^{\text c}_{n}\Big(Q_1=0,Q_2 =-\frac{2}{\beta} \Big) &= \frac{\beta n(n-1)}{2},\\
	\mathfrak{F}^{\text c}_{n}\Big(Q_1=Q_2 = -\frac{2}{\beta}\Big) &= \frac{4}{\beta} - n\Big(2+\frac{\beta}{2}\Big) +\frac{\beta n^2}{2},
\end{split}
\label{vertexneg}
\end{align}  
where the missing vertex follows the symmetry $\mathfrak{F}^{\text c}_{n}(Q_1, Q_2) = \mathfrak{F}^{\text c}_{n}(Q_2, Q_1)$.
It is straightforward to show that for any $n> 1$ and $0<\beta<4$, these vertices are positive and hence, so does $\mathfrak{F}^{\text{c}}_{n}$ in $\mathcal{R}_{Q_1Q_2}^{c}$, for $0<\beta<4$.  The procedure to show $\mathfrak{F}^a_n>0$ for $n\geq1$ is analogous, using  $\mathcal{R}_{Q_1Q_2}^{c}\to \mathcal{R}_{Q_1Q_2}^{a} = [0,4/\beta]\times[0,4/\beta]$. 

\bibliography{refs}

\begin{thebibliography}{49}%
\makeatletter
\providecommand \@ifxundefined [1]{%
 \@ifx{#1\undefined}
}%
\providecommand \@ifnum [1]{%
 \ifnum #1\expandafter \@firstoftwo
 \else \expandafter \@secondoftwo
 \fi
}%
\providecommand \@ifx [1]{%
 \ifx #1\expandafter \@firstoftwo
 \else \expandafter \@secondoftwo
 \fi
}%
\providecommand \natexlab [1]{#1}%
\providecommand \enquote  [1]{``#1''}%
\providecommand \bibnamefont  [1]{#1}%
\providecommand \bibfnamefont [1]{#1}%
\providecommand \citenamefont [1]{#1}%
\providecommand \href@noop [0]{\@secondoftwo}%
\providecommand \href [0]{\begingroup \@sanitize@url \@href}%
\providecommand \@href[1]{\@@startlink{#1}\@@href}%
\providecommand \@@href[1]{\endgroup#1\@@endlink}%
\providecommand \@sanitize@url [0]{\catcode `\\12\catcode `\$12\catcode
  `\&12\catcode `\#12\catcode `\^12\catcode `\_12\catcode `\%12\relax}%
\providecommand \@@startlink[1]{}%
\providecommand \@@endlink[0]{}%
\providecommand \url  [0]{\begingroup\@sanitize@url \@url }%
\providecommand \@url [1]{\endgroup\@href {#1}{\urlprefix }}%
\providecommand \urlprefix  [0]{URL }%
\providecommand \Eprint [0]{\href }%
\providecommand \doibase [0]{https://doi.org/}%
\providecommand \selectlanguage [0]{\@gobble}%
\providecommand \bibinfo  [0]{\@secondoftwo}%
\providecommand \bibfield  [0]{\@secondoftwo}%
\providecommand \translation [1]{[#1]}%
\providecommand \BibitemOpen [0]{}%
\providecommand \bibitemStop [0]{}%
\providecommand \bibitemNoStop [0]{.\EOS\space}%
\providecommand \EOS [0]{\spacefactor3000\relax}%
\providecommand \BibitemShut  [1]{\csname bibitem#1\endcsname}%
\let\auto@bib@innerbib\@empty
\bibitem [{\citenamefont {Verwey}\ and\ \citenamefont
  {Overbeek}(1948)}]{Overbeek}%
  \BibitemOpen
  \bibfield  {author} {\bibinfo {author} {\bibfnamefont {E.~J.~W.}\
  \bibnamefont {Verwey}}\ and\ \bibinfo {author} {\bibfnamefont {J.~T.~G.}\
  \bibnamefont {Overbeek}},\ }\href
  {https://doi.org/https://doi.org/10.1002/pol.1949.120040321} {\emph {\bibinfo
  {title} {“Theory of the stability of lyophobic colloids”}}}\ (\bibinfo
  {publisher} {Elsevier, New York},\ \bibinfo {year} {1948})\BibitemShut
  {NoStop}%
\bibitem [{\citenamefont {Hunter}(2001)}]{hunterbook}%
  \BibitemOpen
  \bibfield  {author} {\bibinfo {author} {\bibfnamefont {R.~J.}\ \bibnamefont
  {Hunter}},\ }\href {https://nla.gov.au/nla.cat-vn1888411} {\emph {\bibinfo
  {title} {Foundations of colloid science}}},\ \bibinfo {edition} {2nd}\ ed.\
  (\bibinfo  {publisher} {Oxford University Press Oxford ; New York},\ \bibinfo
  {year} {2001})\BibitemShut {NoStop}%
\bibitem [{\citenamefont {Jones}(2007)}]{jones2002soft}%
  \BibitemOpen
  \bibfield  {author} {\bibinfo {author} {\bibfnamefont {R.~A.~L.}\
  \bibnamefont {Jones}},\ }\href@noop {} {\emph {\bibinfo {title} {Soft
  Condensed Matter}}}\ (\bibinfo  {publisher} {Oxford University Press},\
  \bibinfo {year} {2007})\BibitemShut {NoStop}%
\bibitem [{\citenamefont {Levin}(1999)}]{LEVIN1999}%
  \BibitemOpen
  \bibfield  {author} {\bibinfo {author} {\bibfnamefont {Y.}~\bibnamefont
  {Levin}},\ }\bibfield  {title} {\bibinfo {title} {When do like charges
  attract?},\ }\href
  {https://doi.org/https://doi.org/10.1016/S0378-4371(98)00552-4} {\bibfield
  {journal} {\bibinfo  {journal} {Physica A: Statistical Mechanics and its
  Applications}\ }\textbf {\bibinfo {volume} {265}},\ \bibinfo {pages} {432}
  (\bibinfo {year} {1999})}\BibitemShut {NoStop}%
\bibitem [{\citenamefont {Trizac}\ and\ \citenamefont
  {\v{S}amaj}(2012)}]{Varenna}%
  \BibitemOpen
  \bibfield  {author} {\bibinfo {author} {\bibfnamefont {E.}~\bibnamefont
  {Trizac}}\ and\ \bibinfo {author} {\bibfnamefont {L.}~\bibnamefont
  {\v{S}amaj}},\ }in\ \href
  {https://doi.org/https://doi.org/10.3254/978-1-61499-278-3-61} {\emph
  {\bibinfo {booktitle} {Proceedings of the International School of Physics
  Enrico Fermi}}},\ Vol.\ \bibinfo {volume} {184},\ \bibinfo {editor} {edited
  by\ \bibinfo {editor} {\bibfnamefont {C.}~\bibnamefont {Bechinger}}, \bibinfo
  {editor} {\bibfnamefont {F.}~\bibnamefont {Sciortino}},\ and\ \bibinfo
  {editor} {\bibfnamefont {P.}~\bibnamefont {Ziherl}}}\ (\bibinfo  {publisher}
  {IOS, Amsterdam},\ \bibinfo {year} {2012})\ pp.\ \bibinfo {pages}
  {61--73}\BibitemShut {NoStop}%
\bibitem [{\citenamefont {Palaia}\ \emph {et~al.}(2020)\citenamefont {Palaia},
  \citenamefont {Telles}, \citenamefont {dos Santos},\ and\ \citenamefont
  {Trizac}}]{Palaia2020}%
  \BibitemOpen
  \bibfield  {author} {\bibinfo {author} {\bibfnamefont {I.}~\bibnamefont
  {Palaia}}, \bibinfo {author} {\bibfnamefont {I.~M.}\ \bibnamefont {Telles}},
  \bibinfo {author} {\bibfnamefont {A.~P.}\ \bibnamefont {dos Santos}},\ and\
  \bibinfo {author} {\bibfnamefont {E.}~\bibnamefont {Trizac}},\ }\bibfield
  {title} {\bibinfo {title} {Electroosmosis as a probe for electrostatic
  correlations},\ }\href {https://doi.org/10.1039/D0SM01523G} {\bibfield
  {journal} {\bibinfo  {journal} {Soft Matter}\ ,\ } (\bibinfo {year}
  {2020})}\BibitemShut {NoStop}%
\bibitem [{\citenamefont {Telles}\ and\ \citenamefont {dos
  Santos}(2021)}]{telles2021}%
  \BibitemOpen
  \bibfield  {author} {\bibinfo {author} {\bibfnamefont {I.~M.}\ \bibnamefont
  {Telles}}\ and\ \bibinfo {author} {\bibfnamefont {A.~P.}\ \bibnamefont {dos
  Santos}},\ }\bibfield  {title} {\bibinfo {title} {Electroosmotic flow grows
  with electrostatic coupling in confining charged dielectric surfaces},\
  }\href {https://doi.org/10.1021/acs.langmuir.0c03116} {\bibfield  {journal}
  {\bibinfo  {journal} {Langmuir}\ }\textbf {\bibinfo {volume} {37}},\ \bibinfo
  {pages} {2104} (\bibinfo {year} {2021})}\BibitemShut {NoStop}%
\bibitem [{\citenamefont {Pellenq}\ and\ \citenamefont
  {Van~Damme}(2004)}]{pellenq2004}%
  \BibitemOpen
  \bibfield  {author} {\bibinfo {author} {\bibfnamefont {R.~J.-M.}\
  \bibnamefont {Pellenq}}\ and\ \bibinfo {author} {\bibfnamefont
  {H.}~\bibnamefont {Van~Damme}},\ }\bibfield  {title} {\bibinfo {title} {Why
  does concrete set?: The nature of cohesion forces in hardened cement-based
  materials},\ }\href {https://doi.org/10.1557/mrs2004.97} {\bibfield
  {journal} {\bibinfo  {journal} {MRS Bulletin}\ }\textbf {\bibinfo {volume}
  {29}},\ \bibinfo {pages} {319–323} (\bibinfo {year} {2004})}\BibitemShut
  {NoStop}%
\bibitem [{\citenamefont {Ioannidou}\ \emph {et~al.}(2016)\citenamefont
  {Ioannidou}, \citenamefont {Kandu{\v{c}}}, \citenamefont {Li}, \citenamefont
  {Frenkel}, \citenamefont {Dobnikar},\ and\ \citenamefont
  {Del~Gado}}]{Ioannidou2016}%
  \BibitemOpen
  \bibfield  {author} {\bibinfo {author} {\bibfnamefont {K.}~\bibnamefont
  {Ioannidou}}, \bibinfo {author} {\bibfnamefont {M.}~\bibnamefont
  {Kandu{\v{c}}}}, \bibinfo {author} {\bibfnamefont {L.}~\bibnamefont {Li}},
  \bibinfo {author} {\bibfnamefont {D.}~\bibnamefont {Frenkel}}, \bibinfo
  {author} {\bibfnamefont {J.}~\bibnamefont {Dobnikar}},\ and\ \bibinfo
  {author} {\bibfnamefont {E.}~\bibnamefont {Del~Gado}},\ }\bibfield  {title}
  {\bibinfo {title} {The crucial effect of early-stage gelation on the
  mechanical properties of cement hydrates},\ }\href
  {https://doi.org/10.1038/ncomms12106} {\bibfield  {journal} {\bibinfo
  {journal} {Nat. Commun.}\ }\textbf {\bibinfo {volume} {7}},\ \bibinfo {pages}
  {12106} (\bibinfo {year} {2016})}\BibitemShut {NoStop}%
\bibitem [{\citenamefont {Holm}\ \emph {et~al.}(2001)\citenamefont {Holm},
  \citenamefont {K{\'e}kicheff},\ and\ \citenamefont {Podgornik}}]{Holm2001}%
  \BibitemOpen
  \bibfield  {author} {\bibinfo {author} {\bibfnamefont {C.}~\bibnamefont
  {Holm}}, \bibinfo {author} {\bibfnamefont {P.}~\bibnamefont
  {K{\'e}kicheff}},\ and\ \bibinfo {author} {\bibfnamefont {R.}~\bibnamefont
  {Podgornik}},\ }\href
  {https://doi.org/https://doi.org/10.1007/978-94-010-0577-7} {\emph {\bibinfo
  {title} {Electrostatic Effects in Soft Matter and Biophysics}}}\ (\bibinfo
  {publisher} {Kluwer Academic, Dordrecht},\ \bibinfo {year}
  {2001})\BibitemShut {NoStop}%
\bibitem [{\citenamefont {Levin}(2002)}]{Levin2002}%
  \BibitemOpen
  \bibfield  {author} {\bibinfo {author} {\bibfnamefont {Y.}~\bibnamefont
  {Levin}},\ }\bibfield  {title} {\bibinfo {title} {Electrostatic correlations:
  from plasma to biology},\ }\href
  {https://doi.org/10.1088/0034-4885/65/11/201} {\bibfield  {journal} {\bibinfo
   {journal} {Rep. Pro. Phys.}\ }\textbf {\bibinfo {volume} {65}},\ \bibinfo
  {pages} {1577} (\bibinfo {year} {2002})}\BibitemShut {NoStop}%
\bibitem [{\citenamefont {Andelman}(2006)}]{andelman2006}%
  \BibitemOpen
  \bibfield  {author} {\bibinfo {author} {\bibfnamefont {D.}~\bibnamefont
  {Andelman}},\ }\bibinfo {title} {Introduction to electrostatics in soft and
  biological matter},\ in\ \href
  {https://books.google.com.co/books?id=hk9UbvyKZE0C} {\emph {\bibinfo
  {booktitle} {Soft Condensed Matter Physics in Molecular and Cell Biology}}},\
  \bibinfo {editor} {edited by\ \bibinfo {editor} {\bibfnamefont {W.~C.~K.}\
  \bibnamefont {Poon}}\ and\ \bibinfo {editor} {\bibfnamefont {D.}~\bibnamefont
  {Andelman}}}\ (\bibinfo  {publisher} {Taylor \& Francis, New York},\ \bibinfo
  {year} {2006})\BibitemShut {NoStop}%
\bibitem [{\citenamefont {Bloomfield}(1996)}]{bloomfield1996}%
  \BibitemOpen
  \bibfield  {author} {\bibinfo {author} {\bibfnamefont {V.~A.}\ \bibnamefont
  {Bloomfield}},\ }\bibfield  {title} {\bibinfo {title} {{DNA} condensation},\
  }\href {https://doi.org/https://doi.org/10.1016/S0959-440X(96)80052-2}
  {\bibfield  {journal} {\bibinfo  {journal} {Current Opinion in Structural
  Biology}\ }\textbf {\bibinfo {volume} {6}},\ \bibinfo {pages} {334} (\bibinfo
  {year} {1996})}\BibitemShut {NoStop}%
\bibitem [{\citenamefont {Caffrey}(2001)}]{Caffrey}%
  \BibitemOpen
  \bibfield  {author} {\bibinfo {author} {\bibfnamefont {M.}~\bibnamefont
  {Caffrey}},\ }\bibfield  {title} {\bibinfo {title} {Structure and dynamic
  properties of membrane lipid and protein},\ }in\ \href
  {https://doi.org/10.1007/978-94-010-0577-7_1} {\emph {\bibinfo {booktitle}
  {Electrostatic Effects in Soft Matter and Biophysics}}},\ \bibinfo {editor}
  {edited by\ \bibinfo {editor} {\bibfnamefont {C.}~\bibnamefont {Holm}},
  \bibinfo {editor} {\bibfnamefont {P.}~\bibnamefont {K{\'e}kicheff}},\ and\
  \bibinfo {editor} {\bibfnamefont {R.}~\bibnamefont {Podgornik}}}\ (\bibinfo
  {publisher} {Kluwer Academic, Dordrecht},\ \bibinfo {year}
  {2001})\BibitemShut {NoStop}%
\bibitem [{\citenamefont {Guldbrand}\ \emph {et~al.}(1984)\citenamefont
  {Guldbrand}, \citenamefont {Jönsson}, \citenamefont {Wennerström},\ and\
  \citenamefont {Linse}}]{Guldbrand1984}%
  \BibitemOpen
  \bibfield  {author} {\bibinfo {author} {\bibfnamefont {L.}~\bibnamefont
  {Guldbrand}}, \bibinfo {author} {\bibfnamefont {B.}~\bibnamefont {Jönsson}},
  \bibinfo {author} {\bibfnamefont {H.}~\bibnamefont {Wennerström}},\ and\
  \bibinfo {author} {\bibfnamefont {P.}~\bibnamefont {Linse}},\ }\bibfield
  {title} {\bibinfo {title} {Electrical double layer forces. a {Monte Carlo}
  study},\ }\href {https://doi.org/https://doi.org/10.1063/1.446912} {\bibfield
   {journal} {\bibinfo  {journal} {J. Chem. Phys.}\ }\textbf {\bibinfo {volume}
  {80}},\ \bibinfo {pages} {2221} (\bibinfo {year} {1984})}\BibitemShut
  {NoStop}%
\bibitem [{\citenamefont {Kjellander}\ and\ \citenamefont
  {Marc\v{e}lja}(1984)}]{Kjellander1984}%
  \BibitemOpen
  \bibfield  {author} {\bibinfo {author} {\bibfnamefont {R.}~\bibnamefont
  {Kjellander}}\ and\ \bibinfo {author} {\bibfnamefont {S.}~\bibnamefont
  {Marc\v{e}lja}},\ }\bibfield  {title} {\bibinfo {title} {Correlation and
  image charge effects in electric double layers},\ }\href
  {https://doi.org/https://doi.org/10.1016/0009-2614(84)87039-6} {\bibfield
  {journal} {\bibinfo  {journal} {Chemical Physics Letters}\ }\textbf {\bibinfo
  {volume} {112}},\ \bibinfo {pages} {49} (\bibinfo {year} {1984})}\BibitemShut
  {NoStop}%
\bibitem [{\citenamefont {Moreira}\ and\ \citenamefont
  {Netz}(2002)}]{Moreira2002MC}%
  \BibitemOpen
  \bibfield  {author} {\bibinfo {author} {\bibfnamefont {A.~G.}\ \bibnamefont
  {Moreira}}\ and\ \bibinfo {author} {\bibfnamefont {R.~R.}\ \bibnamefont
  {Netz}},\ }\bibfield  {title} {\bibinfo {title} {Simulations of counterions
  at charged plates},\ }\href
  {https://doi.org/https://doi.org/10.1140/epje/i2001-10091-9} {\bibfield
  {journal} {\bibinfo  {journal} {Euro. Phys. J. E}\ }\textbf {\bibinfo
  {volume} {8}},\ \bibinfo {pages} {33} (\bibinfo {year} {2002})}\BibitemShut
  {NoStop}%
\bibitem [{\citenamefont {dos Santos}\ and\ \citenamefont
  {Netz}(2018)}]{dosSantos2018}%
  \BibitemOpen
  \bibfield  {author} {\bibinfo {author} {\bibfnamefont {A.~P.}\ \bibnamefont
  {dos Santos}}\ and\ \bibinfo {author} {\bibfnamefont {R.~R.}\ \bibnamefont
  {Netz}},\ }\bibfield  {title} {\bibinfo {title} {Dielectric boundary effects
  on the interaction between planar charged surfaces with counterions only},\
  }\href {https://doi.org/10.1063/1.5022226} {\bibfield  {journal} {\bibinfo
  {journal} {The Journal of Chemical Physics}\ }\textbf {\bibinfo {volume}
  {148}},\ \bibinfo {pages} {164103} (\bibinfo {year} {2018})}\BibitemShut
  {NoStop}%
\bibitem [{\citenamefont {Moreira}\ and\ \citenamefont
  {Netz}(2000)}]{Moreira2000}%
  \BibitemOpen
  \bibfield  {author} {\bibinfo {author} {\bibfnamefont {A.~G.}\ \bibnamefont
  {Moreira}}\ and\ \bibinfo {author} {\bibfnamefont {R.~R.}\ \bibnamefont
  {Netz}},\ }\bibfield  {title} {\bibinfo {title} {Strong-coupling theory for
  counter-ion distributions},\ }\href
  {https://doi.org/10.1209/epl/i2000-00495-1} {\bibfield  {journal} {\bibinfo
  {journal} {Europhysics Letters ({EPL})}\ }\textbf {\bibinfo {volume} {52}},\
  \bibinfo {pages} {705} (\bibinfo {year} {2000})}\BibitemShut {NoStop}%
\bibitem [{\citenamefont {Naji}\ \emph {et~al.}(2005)\citenamefont {Naji},
  \citenamefont {Jungblut}, \citenamefont {Moreira},\ and\ \citenamefont
  {Netz}}]{Naji2005}%
  \BibitemOpen
  \bibfield  {author} {\bibinfo {author} {\bibfnamefont {A.}~\bibnamefont
  {Naji}}, \bibinfo {author} {\bibfnamefont {S.}~\bibnamefont {Jungblut}},
  \bibinfo {author} {\bibfnamefont {A.~G.}\ \bibnamefont {Moreira}},\ and\
  \bibinfo {author} {\bibfnamefont {R.~R.}\ \bibnamefont {Netz}},\ }\bibfield
  {title} {\bibinfo {title} {Electrostatic interactions in strongly coupled
  soft matter},\ }\href
  {https://doi.org/https://doi.org/10.1016/j.physa.2004.12.029} {\bibfield
  {journal} {\bibinfo  {journal} {Physica A}\ }\textbf {\bibinfo {volume}
  {352}},\ \bibinfo {pages} {131 } (\bibinfo {year} {2005})}\BibitemShut
  {NoStop}%
\bibitem [{\citenamefont {{\v{S}}amaj}\ \emph {et~al.}(2018)\citenamefont
  {{\v{S}}amaj}, \citenamefont {Trulsson},\ and\ \citenamefont
  {Trizac}}]{samaj2018}%
  \BibitemOpen
  \bibfield  {author} {\bibinfo {author} {\bibfnamefont {L.}~\bibnamefont
  {{\v{S}}amaj}}, \bibinfo {author} {\bibfnamefont {M.}~\bibnamefont
  {Trulsson}},\ and\ \bibinfo {author} {\bibfnamefont {E.}~\bibnamefont
  {Trizac}},\ }\bibfield  {title} {\bibinfo {title} {Strong-coupling theory of
  counterions between symmetrically charged walls: from crystal to fluid
  phases},\ }\href {https://doi.org/10.1039/C8SM00571K} {\bibfield  {journal}
  {\bibinfo  {journal} {Soft Matter}\ }\textbf {\bibinfo {volume} {14}},\
  \bibinfo {pages} {4040} (\bibinfo {year} {2018})}\BibitemShut {NoStop}%
\bibitem [{\citenamefont {Lieb}(1976)}]{lieb1976}%
  \BibitemOpen
  \bibfield  {author} {\bibinfo {author} {\bibfnamefont {E.~H.}\ \bibnamefont
  {Lieb}},\ }\bibfield  {title} {\bibinfo {title} {The stability of matter},\
  }\href {https://doi.org/10.1103/RevModPhys.48.553} {\bibfield  {journal}
  {\bibinfo  {journal} {Rev. Mod. Phys.}\ }\textbf {\bibinfo {volume} {48}},\
  \bibinfo {pages} {553} (\bibinfo {year} {1976})}\BibitemShut {NoStop}%
\bibitem [{\citenamefont {Trizac}\ and\ \citenamefont
  {T{\'{e}}llez}(2018)}]{Trizac2018}%
  \BibitemOpen
  \bibfield  {author} {\bibinfo {author} {\bibfnamefont {E.}~\bibnamefont
  {Trizac}}\ and\ \bibinfo {author} {\bibfnamefont {G.}~\bibnamefont
  {T{\'{e}}llez}},\ }\bibfield  {title} {\bibinfo {title} {Like-charge
  attraction in a one-dimensional setting: the importance of being odd},\
  }\href {https://doi.org/10.1088/1361-6404/aa9e80} {\bibfield  {journal}
  {\bibinfo  {journal} {European Journal of Physics}\ }\textbf {\bibinfo
  {volume} {39}},\ \bibinfo {pages} {025102} (\bibinfo {year}
  {2018})}\BibitemShut {NoStop}%
\bibitem [{\citenamefont {Lenard}(1961)}]{lenard}%
  \BibitemOpen
  \bibfield  {author} {\bibinfo {author} {\bibfnamefont {A.}~\bibnamefont
  {Lenard}},\ }\bibfield  {title} {\bibinfo {title} {Exact statistical
  mechanics of a one-dimensional system with {Coulomb} forces},\ }\href
  {https://doi.org/http://dx.doi.org/10.1063/1.1703757} {\bibfield  {journal}
  {\bibinfo  {journal} {J. Math. Phys.}\ }\textbf {\bibinfo {volume} {2}},\
  \bibinfo {pages} {682} (\bibinfo {year} {1961})}\BibitemShut {NoStop}%
\bibitem [{\citenamefont {Edwards}\ and\ \citenamefont
  {Lenard}(1962)}]{lenard2}%
  \BibitemOpen
  \bibfield  {author} {\bibinfo {author} {\bibfnamefont {S.~F.}\ \bibnamefont
  {Edwards}}\ and\ \bibinfo {author} {\bibfnamefont {A.}~\bibnamefont
  {Lenard}},\ }\bibfield  {title} {\bibinfo {title} {Exact statistical
  mechanics of a one-dimensional system with {Coulomb} forces. ii. {The} method
  of functional integration},\ }\href
  {https://doi.org/http://dx.doi.org/10.1063/1.1724281} {\bibfield  {journal}
  {\bibinfo  {journal} {J. Math. Phys.}\ }\textbf {\bibinfo {volume} {3}},\
  \bibinfo {pages} {778} (\bibinfo {year} {1962})}\BibitemShut {NoStop}%
\bibitem [{\citenamefont {Prager}(1962)}]{prager}%
  \BibitemOpen
  \bibfield  {author} {\bibinfo {author} {\bibfnamefont {S.}~\bibnamefont
  {Prager}},\ }\bibinfo {title} {The one-dimensional plasma},\ in\ \href
  {https://doi.org/https://doi.org/10.1002/9780470143506.ch5} {\emph {\bibinfo
  {booktitle} {Advances in Chemical Physics}}}\ (\bibinfo  {publisher} {John
  Wiley \& Sons},\ \bibinfo {year} {1962})\ pp.\ \bibinfo {pages}
  {201--224}\BibitemShut {NoStop}%
\bibitem [{\citenamefont {Baxter}(1963)}]{baxter}%
  \BibitemOpen
  \bibfield  {author} {\bibinfo {author} {\bibfnamefont {R.~J.}\ \bibnamefont
  {Baxter}},\ }\bibfield  {title} {\bibinfo {title} {Statistical mechanics of a
  one-dimensional {Coulomb} system with a uniform charge background},\ }\href
  {https://doi.org/10.1017/S0305004100003790} {\bibfield  {journal} {\bibinfo
  {journal} {Math. Proc. Cambridge Philos. Soc.}\ }\textbf {\bibinfo {volume}
  {59}},\ \bibinfo {pages} {779–787} (\bibinfo {year} {1963})}\BibitemShut
  {NoStop}%
\bibitem [{\citenamefont {Dean}\ \emph {et~al.}(2009)\citenamefont {Dean},
  \citenamefont {Horgan},\ and\ \citenamefont {Podgornik}}]{Dean2009}%
  \BibitemOpen
  \bibfield  {author} {\bibinfo {author} {\bibfnamefont {D.~S.}\ \bibnamefont
  {Dean}}, \bibinfo {author} {\bibfnamefont {R.}~\bibnamefont {Horgan}},\ and\
  \bibinfo {author} {\bibfnamefont {R.}~\bibnamefont {Podgornik}},\ }\bibfield
  {title} {\bibinfo {title} {One-dimensional counterion gas between charged
  surfaces: Exact results compared with weak- and strong-coupling analyses},\
  }\href {https://doi.org/https://doi.org/10.1063/1.3078492} {\bibfield
  {journal} {\bibinfo  {journal} {J. Chem. Phys.}\ }\textbf {\bibinfo {volume}
  {130}},\ \bibinfo {pages} {094504} (\bibinfo {year} {2009})}\BibitemShut
  {NoStop}%
\bibitem [{\citenamefont {Frydel}(2019)}]{frydel2019}%
  \BibitemOpen
  \bibfield  {author} {\bibinfo {author} {\bibfnamefont {D.}~\bibnamefont
  {Frydel}},\ }\bibfield  {title} {\bibinfo {title} {One-dimensional {Coulomb}
  system in a sticky wall confinement: Exact results},\ }\href
  {https://doi.org/https://doi.org/10.1103/PhysRevE.100.042113} {\bibfield
  {journal} {\bibinfo  {journal} {Phys. Rev. E}\ }\textbf {\bibinfo {volume}
  {100}},\ \bibinfo {pages} {042113} (\bibinfo {year} {2019})}\BibitemShut
  {NoStop}%
\bibitem [{\citenamefont {Jancovici}(1981)}]{Jancovici1981}%
  \BibitemOpen
  \bibfield  {author} {\bibinfo {author} {\bibfnamefont {B.}~\bibnamefont
  {Jancovici}},\ }\bibfield  {title} {\bibinfo {title} {Exact results for the
  two-dimensional one-component plasma},\ }\href
  {https://doi.org/10.1103/PhysRevLett.46.386} {\bibfield  {journal} {\bibinfo
  {journal} {Phys. Rev. Lett.}\ }\textbf {\bibinfo {volume} {46}},\ \bibinfo
  {pages} {386} (\bibinfo {year} {1981})}\BibitemShut {NoStop}%
\bibitem [{\citenamefont {Hansen}\ and\ \citenamefont
  {Viot}(1985)}]{Hansen1985}%
  \BibitemOpen
  \bibfield  {author} {\bibinfo {author} {\bibfnamefont {J.~P.}\ \bibnamefont
  {Hansen}}\ and\ \bibinfo {author} {\bibfnamefont {P.}~\bibnamefont {Viot}},\
  }\bibfield  {title} {\bibinfo {title} {Two-body correlations and pair
  formation in the two-dimensional {C}oulomb gas},\ }\href
  {https://doi.org/10.1007/BF01010417} {\bibfield  {journal} {\bibinfo
  {journal} {Journal of Statistical Physics}\ }\textbf {\bibinfo {volume}
  {38}},\ \bibinfo {pages} {823} (\bibinfo {year} {1985})}\BibitemShut
  {NoStop}%
\bibitem [{\citenamefont {Minnhagen}(1987)}]{Minnhagen_1987}%
  \BibitemOpen
  \bibfield  {author} {\bibinfo {author} {\bibfnamefont {P.}~\bibnamefont
  {Minnhagen}},\ }\bibfield  {title} {\bibinfo {title} {The two-dimensional
  {C}oulomb gas, vortex unbinding, and superfluid-superconducting films},\
  }\href {https://doi.org/10.1103/RevModPhys.59.1001} {\bibfield  {journal}
  {\bibinfo  {journal} {Rev. Mod. Phys.}\ }\textbf {\bibinfo {volume} {59}},\
  \bibinfo {pages} {1001} (\bibinfo {year} {1987})}\BibitemShut {NoStop}%
\bibitem [{\citenamefont {Cornu}\ and\ \citenamefont
  {Jancovici}(1989)}]{cornu1989}%
  \BibitemOpen
  \bibfield  {author} {\bibinfo {author} {\bibfnamefont {F.}~\bibnamefont
  {Cornu}}\ and\ \bibinfo {author} {\bibfnamefont {B.}~\bibnamefont
  {Jancovici}},\ }\bibfield  {title} {\bibinfo {title} {The electrical double
  layer: {A} solvable model},\ }\href {https://doi.org/10.1063/1.455986}
  {\bibfield  {journal} {\bibinfo  {journal} {The Journal of Chemical Physics}\
  }\textbf {\bibinfo {volume} {90}},\ \bibinfo {pages} {2444} (\bibinfo {year}
  {1989})}\BibitemShut {NoStop}%
\bibitem [{\citenamefont {{\v{S}}amaj}(2003)}]{samaj2003}%
  \BibitemOpen
  \bibfield  {author} {\bibinfo {author} {\bibfnamefont {L.}~\bibnamefont
  {{\v{S}}amaj}},\ }\bibfield  {title} {\bibinfo {title} {Exact solution of a
  charge-asymmetric two-dimensional {C}oulomb gas},\ }\href
  {https://doi.org/10.1023/A:1022209108732} {\bibfield  {journal} {\bibinfo
  {journal} {Journal of Statistical Physics}\ }\textbf {\bibinfo {volume}
  {111}},\ \bibinfo {pages} {261} (\bibinfo {year} {2003})}\BibitemShut
  {NoStop}%
\bibitem [{\citenamefont {{\v{S}}amaj}\ and\ \citenamefont
  {Trav{\v{e}}nec}(2000)}]{samaj2000}%
  \BibitemOpen
  \bibfield  {author} {\bibinfo {author} {\bibfnamefont {L.}~\bibnamefont
  {{\v{S}}amaj}}\ and\ \bibinfo {author} {\bibfnamefont {I.}~\bibnamefont
  {Trav{\v{e}}nec}},\ }\bibfield  {title} {\bibinfo {title} {Thermodynamic
  properties of the two-dimensional two-component plasma},\ }\href
  {https://doi.org/10.1023/A:1026489924895} {\bibfield  {journal} {\bibinfo
  {journal} {Journal of Statistical Physics}\ }\textbf {\bibinfo {volume}
  {101}},\ \bibinfo {pages} {713} (\bibinfo {year} {2000})}\BibitemShut
  {NoStop}%
\bibitem [{\citenamefont {Coleman}(1975)}]{coleman1975}%
  \BibitemOpen
  \bibfield  {author} {\bibinfo {author} {\bibfnamefont {S.}~\bibnamefont
  {Coleman}},\ }\bibfield  {title} {\bibinfo {title} {Quantum sine-{Gordon}
  equation as the massive {Thirring} model},\ }\href
  {https://doi.org/10.1103/PhysRevD.11.2088} {\bibfield  {journal} {\bibinfo
  {journal} {Phys. Rev. D}\ }\textbf {\bibinfo {volume} {11}},\ \bibinfo
  {pages} {2088} (\bibinfo {year} {1975})}\BibitemShut {NoStop}%
\bibitem [{\citenamefont {T\'ellez}(2005)}]{tellez_2005}%
  \BibitemOpen
  \bibfield  {author} {\bibinfo {author} {\bibfnamefont {G.}~\bibnamefont
  {T\'ellez}},\ }\bibfield  {title} {\bibinfo {title} {Short-distance expansion
  of correlation functions for charge-symmetric two-dimensional two-component
  plasma: exact results},\ }\href
  {https://doi.org/10.1088/1742-5468/2005/10/p10001} {\bibfield  {journal}
  {\bibinfo  {journal} {Journal of Statistical Mechanics: Theory and
  Experiment}\ }\textbf {\bibinfo {volume} {2005}},\ \bibinfo {pages} {P10001}
  (\bibinfo {year} {2005})}\BibitemShut {NoStop}%
\bibitem [{\citenamefont {\v{S}amaj}(2005)}]{samaj2005}%
  \BibitemOpen
  \bibfield  {author} {\bibinfo {author} {\bibfnamefont {L.}~\bibnamefont
  {\v{S}amaj}},\ }\bibfield  {title} {\bibinfo {title} {Anomalous effects of
  ``guest'' charges immersed in electrolyte: Exact {2D} results},\ }\href
  {https://doi.org/10.1007/s10955-005-5477-8} {\bibfield  {journal} {\bibinfo
  {journal} {Journal of Statistical Physics}\ }\textbf {\bibinfo {volume}
  {120}},\ \bibinfo {pages} {125} (\bibinfo {year} {2005})}\BibitemShut
  {NoStop}%
\bibitem [{\citenamefont {T{\'e}llez}(2006)}]{Tellez2006Jstat}%
  \BibitemOpen
  \bibfield  {author} {\bibinfo {author} {\bibfnamefont {G.}~\bibnamefont
  {T{\'e}llez}},\ }\bibfield  {title} {\bibinfo {title} {Guest charges in an
  electrolyte: Renormalized charge, long- and short-distance behavior of the
  electric potential and density profiles},\ }\href
  {https://doi.org/10.1007/s10955-005-8069-8} {\bibfield  {journal} {\bibinfo
  {journal} {Journal of Statistical Physics}\ }\textbf {\bibinfo {volume}
  {122}},\ \bibinfo {pages} {787} (\bibinfo {year} {2006})}\BibitemShut
  {NoStop}%
\bibitem [{\citenamefont {{\v{S}}amaj}(2006)}]{samaj2006}%
  \BibitemOpen
  \bibfield  {author} {\bibinfo {author} {\bibfnamefont {L.}~\bibnamefont
  {{\v{S}}amaj}},\ }\bibfield  {title} {\bibinfo {title} {Renormalization of a
  hard-core guest charge immersed in a two-dimensional electrolyte},\ }\href
  {https://doi.org/10.1007/s10955-006-9122-y} {\bibfield  {journal} {\bibinfo
  {journal} {Journal of Statistical Physics}\ }\textbf {\bibinfo {volume}
  {124}},\ \bibinfo {pages} {1179} (\bibinfo {year} {2006})}\BibitemShut
  {NoStop}%
\bibitem [{\citenamefont {T\'ellez}(2006)}]{tellez2006EPL}%
  \BibitemOpen
  \bibfield  {author} {\bibinfo {author} {\bibfnamefont {G.}~\bibnamefont
  {T\'ellez}},\ }\bibfield  {title} {\bibinfo {title} {Charge inversion of
  colloids in an exactly solvable model},\ }\href
  {https://doi.org/10.1209/epl/i2006-10389-8} {\bibfield  {journal} {\bibinfo
  {journal} {Europhys. Lett.}\ }\textbf {\bibinfo {volume} {76}},\ \bibinfo
  {pages} {1186} (\bibinfo {year} {2006})}\BibitemShut {NoStop}%
\bibitem [{\citenamefont {Fateev}\ \emph {et~al.}(1998)\citenamefont {Fateev},
  \citenamefont {Lukyanov}, \citenamefont {Zamolodchikov},\ and\ \citenamefont
  {Zamolodchikov}}]{FATEEV1998}%
  \BibitemOpen
  \bibfield  {author} {\bibinfo {author} {\bibfnamefont {V.}~\bibnamefont
  {Fateev}}, \bibinfo {author} {\bibfnamefont {S.}~\bibnamefont {Lukyanov}},
  \bibinfo {author} {\bibfnamefont {A.}~\bibnamefont {Zamolodchikov}},\ and\
  \bibinfo {author} {\bibfnamefont {A.}~\bibnamefont {Zamolodchikov}},\
  }\bibfield  {title} {\bibinfo {title} {Expectation values of local fields in
  the {Bullough-Dodd} model and integrable perturbed conformal field
  theories},\ }\href
  {https://doi.org/https://doi.org/10.1016/S0550-3213(98)00002-9} {\bibfield
  {journal} {\bibinfo  {journal} {Nuclear Physics B}\ }\textbf {\bibinfo
  {volume} {516}},\ \bibinfo {pages} {652} (\bibinfo {year}
  {1998})}\BibitemShut {NoStop}%
\bibitem [{\citenamefont {Baseilhac}\ and\ \citenamefont
  {Stanishkov}(2001)}]{BASEILHAC}%
  \BibitemOpen
  \bibfield  {author} {\bibinfo {author} {\bibfnamefont {P.}~\bibnamefont
  {Baseilhac}}\ and\ \bibinfo {author} {\bibfnamefont {M.}~\bibnamefont
  {Stanishkov}},\ }\bibfield  {title} {\bibinfo {title} {Expectation values of
  descendent fields in the {Bullough–Dodd} model and related perturbed
  conformal field theories},\ }\href
  {https://doi.org/https://doi.org/10.1016/S0550-3213(01)00287-5} {\bibfield
  {journal} {\bibinfo  {journal} {Nuclear Physics B}\ }\textbf {\bibinfo
  {volume} {612}},\ \bibinfo {pages} {373 } (\bibinfo {year}
  {2001})}\BibitemShut {NoStop}%
\bibitem [{\citenamefont {Dotsenko}\ and\ \citenamefont
  {Fateev}(1984)}]{DOTSENKO}%
  \BibitemOpen
  \bibfield  {author} {\bibinfo {author} {\bibfnamefont {V.}~\bibnamefont
  {Dotsenko}}\ and\ \bibinfo {author} {\bibfnamefont {V.}~\bibnamefont
  {Fateev}},\ }\bibfield  {title} {\bibinfo {title} {Conformal algebra and
  multipoint correlation functions in {2D} statistical models},\ }\href
  {https://doi.org/https://doi.org/10.1016/0550-3213(84)90269-4} {\bibfield
  {journal} {\bibinfo  {journal} {Nuclear Physics B}\ }\textbf {\bibinfo
  {volume} {240}},\ \bibinfo {pages} {312 } (\bibinfo {year}
  {1984})}\BibitemShut {NoStop}%
\bibitem [{\citenamefont {Dotsenko}\ and\ \citenamefont
  {Fateev}(1985)}]{DOTSENKO2}%
  \BibitemOpen
  \bibfield  {author} {\bibinfo {author} {\bibfnamefont {V.~S.}\ \bibnamefont
  {Dotsenko}}\ and\ \bibinfo {author} {\bibfnamefont {V.~A.}\ \bibnamefont
  {Fateev}},\ }\bibfield  {title} {\bibinfo {title} {Four-point correlation
  functions and the operator algebra in {2D} conformal invariant theories with
  central charge {$C\leq 1$}},\ }\href
  {https://doi.org/10.1016/S0550-3213(85)80004-3} {\bibfield  {journal}
  {\bibinfo  {journal} {Nuclear Physics B}\ }\textbf {\bibinfo {volume}
  {251}},\ \bibinfo {pages} {691} (\bibinfo {year} {1985})}\BibitemShut
  {NoStop}%
\bibitem [{\citenamefont {Aomoto}(1987)}]{AOMOTO}%
  \BibitemOpen
  \bibfield  {author} {\bibinfo {author} {\bibfnamefont {K.}~\bibnamefont
  {Aomoto}},\ }\bibfield  {title} {\bibinfo {title} {On the complex {S}elberg
  integral},\ }\href {https://doi.org/10.1093/qmath/38.4.385} {\bibfield
  {journal} {\bibinfo  {journal} {The Quarterly Journal of Mathematics}\
  }\textbf {\bibinfo {volume} {38}},\ \bibinfo {pages} {385} (\bibinfo {year}
  {1987})}\BibitemShut {NoStop}%
\bibitem [{\citenamefont {Smirnov}(1992)}]{Smirnov1992}%
  \BibitemOpen
  \bibfield  {author} {\bibinfo {author} {\bibfnamefont {F.~A.}\ \bibnamefont
  {Smirnov}},\ }\href {https://doi.org/https://doi.org/10.1142/1115} {\emph
  {\bibinfo {title} {Form Factors in Completely Integrable Models of Quantum
  Field Theory}}}\ (\bibinfo  {publisher} {World Scientific, Singapore},\
  \bibinfo {year} {1992})\BibitemShut {NoStop}%
\bibitem [{bet()}]{beta8d3}%
  \BibitemOpen
  \href@noop {} {\bibinfo {title} {The large-distance behavior for large
  coulombic couplings ($\beta\geq8/3$) has not been computed, to the authors
  knowledge.}}\BibitemShut {Stop}%
\bibitem [{nor()}]{normG}%
  \BibitemOpen
  \href@noop {} {\bibinfo {title} {{The normalization of the large-distance
  interaction strength is such that $\max \{ \, | \, \widetilde{
  \mathcal{G}^{\infty}_{\text{eff}} }(Q_1,Q_2;\beta) \, | \, \} = \max \{ \, |
  \, \mathcal{G}_{\text{eff}}(Q_1,Q_2;\beta) \, | \, \} $. This allows to
  compare the sign of these quantities, which has the same meaning in both
  asymptotic cases at variance to the magnitudes . }}}\BibitemShut {NoStop}%
\end{thebibliography}%

\end{document}